\documentclass[11pt]{article}
\usepackage{geometry,amsmath,amssymb,graphicx,stmaryrd,enumerate,theorem}
\usepackage[pdftex,colorlinks=true,linkcolor=blue,extension=pdf]{hyperref}
\newcommand{\cdummy}{\cdot}
\newcommand{\matheuler}{\gamma}
\newcommand{\nobracket}{}
\newcommand{\tmaffiliation}[1]{\\ #1}
\newcommand{\tmcolor}[2]{{\color{#1}{#2}}}
\newcommand{\tmdummy}{$\mbox{}$}
\newcommand{\tmem}[1]{{\em #1\/}}
\newcommand{\tmemail}[1]{\\ \textit{Email:} \texttt{#1}}
\newcommand{\tmmathbf}[1]{\ensuremath{\boldsymbol{#1}}}
\newcommand{\tmname}[1]{\textsc{#1}}
\newcommand{\tmop}[1]{\ensuremath{\operatorname{#1}}}
\newcommand{\tmscript}[1]{\text{\scriptsize{$#1$}}}
\newcommand{\tmstrong}[1]{\textbf{#1}}
\newcommand{\tmtextit}[1]{{\itshape{#1}}}
\newcommand{\tmtextsc}[1]{{\scshape{#1}}}
\newcommand{\tmtextsf}[1]{{\sffamily{#1}}}
\newcommand{\tmverbatim}[1]{{\ttfamily{#1}}}
\newenvironment{descriptioncompact}{\begin{description} }{\end{description}}
\newenvironment{enumeratenumeric}{\begin{enumerate}[1.] }{\end{enumerate}}
\newenvironment{quoteenv}{\begin{quote} }{\end{quote}}
\newtheorem{definition}{Definition}
{\theorembodyfont{\rmfamily}\newtheorem{example}{Example}}
\newtheorem{lemma}{Lemma}
\newtheorem{proposition}{Proposition}
{\theorembodyfont{\rmfamily}\newtheorem{remark}{Remark}}
\newtheorem{theorem}{Theorem}
\newcommand{\tmkeywords}{\textbf{Keywords:} }

\numberwithin{equation}{section}
\numberwithin{theorem}{section}
\numberwithin{lemma}{section}
\numberwithin{proposition}{section}
\numberwithin{figure}{section}
\numberwithin{table}{section}
\numberwithin{example}{section}
\numberwithin{definition}{section}
\numberwithin{footnote}{section}
\numberwithin{remark}{section}

\newcommand{\maxent}{\tmtextsc{MaxEnt}}
\newcommand{\maxgent}{\tmtextsc{MaxGEnt}}

\begin{document}

\title{Generalized Entropy Concentration for Counts}

\author{
  Kostas N. Oikonomou
  \tmaffiliation{AT\&T Labs Research\\
  Middletown, NJ 07748, U.S.A.}
  \tmemail{ko@research.att.com}
}

\date{May 2019}

\maketitle

\begin{abstract}
  The phenomenon of entropy concentration provides strong support for the
  maximum entropy method, {\maxent}, for inferring a probability vector from
  information in the form of constraints. Here we extend this phenomenon, in a
  discrete setting, to non-negative integral vectors not necessarily summing
  to 1. We show that linear constraints that simply bound the allowable sums
  suffice for concentration to occur even in this setting. This requires a
  new, `generalized' entropy measure in which the sum of the vector plays a
  role. We measure the concentration in terms of deviation from the maximum
  generalized entropy value, or in terms of the distance from the maximum
  generalized entropy vector. We provide non-asymptotic bounds on the
  concentration in terms of various parameters, including a tolerance on the
  constraints which ensures that they are always satisfied by an integral
  vector. Generalized entropy maximization is not only compatible with
  ordinary {\maxent}, but can also be considered an extension of it, as it
  allows us to address problems that cannot be formulated as {\maxent}
  problems.
\end{abstract}

\tmkeywords{maximum generalized entropy, counts, concentration, linear
constraints, inequalities, norms, tolerances}

{\tableofcontents}

\section{Introduction\label{sec:intro}}

The maximum entropy method or principle, originally proposed by E.T. Jaynes in
1957, now appears in standard textbooks on engineering probability and
information theory, {\cite{PP2002}}, {\cite{CT2}}. Commonly referred to as
{\maxent}, the principle essentially states that if the only information
available about a probability vector is in the form of linear constraints on
its elements, then, among all others, the preferred probability vector is the
one that maximizes the Shannon entropy under these constraints. Besides the
great wealth and diversity of its applications, {\maxent} can be justified on
a variety of theoretical grounds: axiomatic formulations ({\cite{SJ1980}},
{\cite{Skilling1989}}, {\cite{Csi1991}}, {\cite{CatEIFP}}), the concentration
phenomenon ({\cite{JP}}, {\cite{Gr2008}}, {\cite{CatEIFP}},
{\cite{entc2016}}), decision- and game-theoretic interpretations
({\cite{Gr2008}} and references therein), and its unification with Bayesian
inference ({\cite{Giffin2007}}, {\cite{CatEIFP}}).

Among these justifications, in a discrete setting, the appeal of concentration
lies in its conceptual simplicity. It is essentially a combinatorial argument,
first presented by E.T. Jaynes {\cite{JP}}, who called it ``concentration of
distributions at entropy maxima''. The concentration viewpoint was further
developed in {\cite{Gr2008}} and {\cite{entc2016}}, which presented
generalizations, improved results, eliminated the asymptotics, and studied
additional aspects.

In this paper we adopt a discrete, finite, non-probabilistic, combinatorial
approach, and show that the concentration phenomenon arises in a new setting,
that of non-negative vectors which are not necessarily density
vectors\footnote{What we call here `density' or `frequency' vectors would be
  called ``discrete probability distributions'', possibly `empirical', if we
  were operating in a probabilistic setting.}. Among other things, this requires
introducing a new, `generalized' entropy measure.  This new concentration
phenomenon lends support to an extension of the {\maxent} method to what we call
``maximum generalized entropy'', or {\maxgent}.

The basics of entropy concentration are easiest to explain in terms of the
abstract ``balls and bins'' paradigm ({\cite{JL}}). There are $m$ labelled,
distinguishable bins, to which $n$ indistinguishable balls are to be allocated
one-by-one. The final content of the bins is described by a count vector $\nu
= (\nu_1, \ldots, \nu_m)$ which sums to $n$, and a corresponding frequency
vector $f = \nu / n$, summing to 1. Suppose that the frequency vector must
satisfy a set of linear equalities and inequalities, $\sum_i a_{i j} f_i =
b_j$ and $\sum_i a_{i j} f_i \leqslant b_j$, with $a_{i j}, b_j \in
\mathbb{R}$. The {\tmem{concentration phenomenon}} is that as $n$ becomes
large, the overwhelming majority of the allocations which accord with the
constraints have frequency vectors that are close to the $m$-vector which
maximizes the Shannon entropy subject to the constraints.

In our extension there is no longer a given number of balls. Therefore we
cannot define a unique frequeny vector, but must deal directly with count
vectors $\nu$ whose sums are unknown (Example \ref{ex:first} below makes this
clear). The linear constraints are now placed on the counts $\nu_i$, again
with coefficients in $\mathbb{R}$. Our only assumption about the constraints
is that they limit the \tmtextit{sums} of the count vectors to lie in a finite
range $[s_1, s_2]$. With just this assumption, we show that as the counts are
allowed to become larger and larger (by a process of {\tmem{scaling}} the
problem, explained in {\textsection}\ref{sec:basic}), the vast majority of
allocations that satisfy the constraints in fact have count vectors close to
the non-negative $m$-vector $x^{\ast}$ that maximizes the {\tmem{generalized
entropy}} $G (x)$. A precise statement of this concentration phenomenon needs
some additional preliminaries, and is given at the end of this section.

Our main results are, in {\textsection}\ref{sec:G}, a new generalized entropy
function $G$, defined on arbitrary non-negative vectors, which reduces to the
Shannon entropy $H$ on vectors summing to 1; its properties are studied in
{\textsection}\ref{sec:G} and {\textsection}\ref{sec:basic}, where the scaling
process is also introduced. In {\textsection}\ref{sec:gconc} we demonstrate
the new concentration pheonomenon with respect to deviations from the maximum
generalized entropy value $G^{\ast}$. Theorem \ref{th:Gdiff} gives a lower
bound on the ratio of the number of realizations of the {\maxgent} vector to
that of the \tmtextit{set} of count vectors $\nu$ whose generalized entropies
$G (\nu)$ are far from the maximum {\tmem{value}} $G^{\ast}$. Then Theorem
\ref{th:cGdiff} completes the picture by deriving how large the problem must
be for the above ratio to be suitably large. In {\textsection}\ref{sec:dconc}
we establish concentration with respect to the $\ell_1$ norm
\tmtextit{distance} of the count vectors from the {\maxgent} {\tmem{vector}};
we present Theorems \ref{th:cdist1} and \ref{th:cdist2}, which are analogous
to those of {\textsection}\ref{sec:gconc}, and also Theorem \ref{th:cdist3},
an optimized version of Theorem \ref{th:cdist2}. In all the theorems, `far',
`large', etc. are defined in terms of parameters, introduced in Table
\ref{tab:t} below. None of our results involve any asymptotic considerations,
and we give a number of numerical illustrations.

The following example demonstrates the basic issues referred to above in a
very simple setting, which highlights the differences with the usual frequency
vector case. After this, we proceed to the precise statement of generalized
entropy concentration.

\begin{example}
  \label{ex:first}A number of indistinguishable balls\footnote{The balls don't
  {\tmem{have}} to be indistinguishable, we just ignore distinguishing
  characteristics, if they have any. However, in modelling some situations,
  such as in Example \ref{ex:imp}, indistinguishability is essential.} are to
  be placed one-by-one in three bins, red, green, and blue. The final content
  $(\nu_r, \nu_g, \nu_b)$ of the bins must satisfy $\nu_r + \nu_g = 4$ and
  $\nu_g + \nu_b \leqslant 6$. Thus the total number of balls that may be put
  in the bins cannot be too small, e.g. 3, or too large, e.g. 20. Each
  assignment of balls to the bins is described by a sequence made from the
  letters $r, g, b$, with a corresponding {\tmem{count}} vector $\nu = (\nu_r,
  \nu_g, \nu_b)$; the sequence can be of {\tmem{any}} length $n$ consistent
  with the constraints. Table \ref{tab:ass} lists all the count vectors that
  satisfy the constraints, their sums $n$, and their number of
  {\tmem{realizations}} $\# \nu$, i.e. the number of sequences that result in
  these counts, given by a multinomial coefficient, e.g. $\binom{7}{3, 1, 3} =
  140$. [In the terminology of the theory of types, $\# \nu$ is the size of
  the type class $T (\nu / n)$.] What can be said about the ``most likely''
  final content of the bins?
  
  \begin{table}[h]
    \centering
    \begin{tabular}{l}
      \begin{tabular}{|lll|l|l|} \hline
        $\nu_r$ & $\nu_g$ & $\nu_b$ & $n$ & $\# \nu$\\
        \hline
        0 & 4 & 0 & 4 & 1\\
        1 & 3 & 0 &  & 4\\
        2 & 2 & 0 &  & 6\\
        3 & 1 & 0 &  & 4\\
        4 & 0 & 1 &  & 1\\
        \hline
        0 & 4 & 1 & 5 & 5\\
        1 & 3 & 1 &  & 20\\
        2 & 2 & 1 &  & 30\\
        3 & 1 & 1 &  & 20\\
        4 & 0 & 1 &  & 5 \\ \hline
      \end{tabular}
    \end{tabular}\quad\begin{tabular}{l}
      \begin{tabular}{|lll|l|l|} \hline
        $\nu_r$ & $\nu_g$ & $\nu_b$ & $n$ & $\# \nu$\\
        \hline
        0 & 4 & 2 & 6 & 15\\
        1 & 3 & 2 &  & 60\\
        2 & 2 & 2 &  & 90\\
        3 & 1 & 2 &  & 60\\
        4 & 0 & 2 &  & 15\\
        \hline
        1 & 3 & 3 & 7 & 140\\
        2 & 2 & 3 &  & 210\\
        3 & 1 & 3 &  & 140\\
        4 & 0 & 3 &  & 35 \\ \hline
      \end{tabular}
    \end{tabular}\quad\begin{tabular}{l}
      \begin{tabular}{|lll|l|l|} \hline
        $\nu_r$ & $\nu_g$ & $\nu_b$ & $n$ & $\# \nu$\\
        \hline
        2 & 2 & 4 & 8 & 420\\
        3 & 1 & 4 &  & 280\\
        4 & 0 & 4 &  & 70\\
        \hline
        {\tmstrong{3}} & {\tmstrong{1}} & {\tmstrong{5}} & {\tmstrong{9}} &
        {\tmstrong{504}}\\
        4 & 0 & 5 &  & 126\\
        \hline
        4 & 0 & 6 & 10 & 210 \\ \hline
      \end{tabular}
    \end{tabular}
    \caption{\label{tab:ass}\small The count vectors $\nu = (\nu_r, \nu_g,
      \nu_b)$ satisfying $\nu_r + \nu_g = 4, \nu_g + \nu_b \leqslant 6$, their
      sum $n$, and their number of realizations $\# \nu$. If we had the
      additional constraint $\nu_r + \nu_g + \nu_b = 7$, only the $n = 7$
      section of the table would apply, and we would reduce to a {\maxent}
      problem.}
  \end{table}
  
  This example makes two points. First, it does not seem possible to find a
  single {\tmem{frequency}} vector that can be naturally associated with the
  problem; without that, one cannot think about maximizing the usual
  entropy\footnote{In the ordinary entropy problem where we have a single $n$,
  the distinction between {\tmem{count}} and {\tmem{frequency}} vectors
  doesn't really matter, there is a 1-1 correspondence; but this is not true
  here.}. Second, one may think that starting with the {\tmem{largest}}
  possible number of balls, 10 in this case, would lead to the greatest number
  of realizations. But this is not so: the count vector with the most
  realizations sums to 9, and even vectors summing to 8 have more realizations
  than the one summing to 10.
\end{example}

Next we give a precise statement, \tmtextit{GC} below, of generalized entropy
concentration. To do that we need to (a) define the generalized entropy and
describe how to find the vector that maximizes it, (b) specify how to derive
the bounds $s_1, s_2$ from the constraints, (c) describe how to ensure the
existence of integral solutions (count vectors) to the constraints, and (d)
introduce parameters that define the concentration.

To find the vector with the largest number of realizations in a problem like
that of Example \ref{ex:first}, we first assume that the problem {\tmem{does
not admit arbitrarily large solutions}}. This is made precise in
(\ref{eq:s1s2}) below, but a necessary condition is that each element of $\nu$
appears in some constraint\footnote{But this is not sufficient: consider, e.g.
$m = 2$ and $\nu_1, \nu_2 \geqslant 0$, $\nu_1 - \nu_2 = 10$.}. Next we relax
the integrality requirement on the counts, and set up a continuous
maximization problem
\begin{equation}
  \begin{array}{c}
    \max_{x \in \mathcal{C}} G (x), \text{\quad where} \quad G (x) = - \sum_i
    x_i \ln x_i + \left( \sum_i x_i \right) \ln \left( \sum_i x_i \right)\\
    \text{and} \qquad \mathcal{C}= \{ x \in \mathbb{R}^m \mid A^E x = b^E,
    A^I x \leqslant b^I, x \geqslant 0 \} .
  \end{array} \label{eq:maxG},
\end{equation}
Here $G (x)$ is the \tmtextit{generalized entropy} of the real vector $x
\geqslant 0$, and the constraints on $x$ are expressed via the real matrices
$A^E, A^I$ and vectors $b^E, b^I$. We assume that the constraints (a) are
satisfiable, and (b) they {\tmem{bound}} the possible {\tmem{sums}} of the $x
\in \mathcal{C}$; this is equivalent to assuming that all $x_i$ are bounded.
Thus $\mathcal{C}$ is a non-empty polytope in $\mathbb{R}^m$ and
(\ref{eq:maxG}) is a concave maximization problem (see e.g. {\cite{BV}}) with
a solution $x^{\ast}$. We will refer to (\ref{eq:maxG}) as the ``{\maxgent}
problem'' and to $x^{\ast}$ as ``the {\maxgent} vector'' or as ``the optimal
relaxed count vector''. Since the function $G$ is concave but not strictly
concave, see Fig. \ref{fig:G2} in {\textsection}\ref{sec:G}, it is not
immediate that the solution $x^{\ast}$ is unique; however, we show that this
is the case in {\textsection}\ref{sec:max}.

The boundedness assumption is that $\sum_i x_i$ lies between (finite) numbers
$s_1$ and $s_2$; these are determined by solving the linear programs
\begin{equation}
  s_1 \; \triangleq \min_{x \in \mathcal{C}}  (x_1 + \cdots + x_m), \qquad s_2
  \; \triangleq \max_{x \in \mathcal{C}}  (x_1 + \cdots + x_m) .
  \label{eq:s1s2}
\end{equation}
(A technicality is that the constraints may force some elements of $x^{\ast}$
to be 0; for reasons explained in {\textsection}\ref{sec:basic} it is
convenient to eliminate such elements, so that in the end all elements of
$x^{\ast}$ can be assumed to be positive reals.) Finally, from $x^{\ast}$ we
derive an \tmtextit{integral} vector $\nu^{\ast}$, to which we refer as the
optimal, or {\maxgent} {\tmem{count}} vector, by a procedure explained in
{\textsection}\ref{sec:basic}.

Because in the end we are interested only in integral/count vectors in the set
$\mathcal{C}$ of (\ref{eq:maxG}), we will introduce, as explained in
{\textsection}\ref{sec:basic}, \tmtextit{tolerances} on the satisfaction on
the constraints, governed by a parameter $\delta$. This will turn
$\mathcal{C}$ into $\mathcal{C} (\delta)$. To describe the concentration we
need two more parameters, $\varepsilon$ specifying the {\tmem{strength}} of
the concentration, and $\eta$ or $\vartheta$ describing the {\tmem{size of the
region}} in which it occurs. The parameters are summarized in Table
\ref{tab:t}.

\begin{table}[h]
  \centering
  \begin{tabular}{|ll|} \hline
    $\delta$: & relative tolerance in satisfying the constraints\\
    $\varepsilon$: & concentration tolerance, on number of realizations\\
    $\eta$: & relative tolerance in deviation from the maximum generalized
    entropy value $G^{\ast}$\\
    $\vartheta$: & absolute tolerance in deviation (distance) from the optimal
    relaxed count vector $x^{\ast}$ \\ \hline
  \end{tabular}
  \caption{\label{tab:t}\small Parameters for the concentration results.}
\end{table}

Lastly, when we have ordinary entropy and frequency vectors, concentration
occurs by increasing the number of balls $n$. With count vectors, this is
replaced by increasing $b^E, b^I$, the values of the constraints. The increase
we consider here consists in multiplying these vectors by a scalar $c > 1$, a
process which we call \tmtextit{scaling}. This scaling results in larger and
larger count vectors being admissible and is described in detail in
{\textsection}\ref{sec:basic}.

Now we can give the precise statement of the concentration phenomenon for
count vectors:

\begin{quoteenv}
  \tmtextit{GC}: Theorems \ref{th:cGdiff} and \ref{th:cdist2} compute a number
  $\hat{c} (\delta, \varepsilon, \eta)$ and $\hat{c} (\delta, \varepsilon,
  \vartheta)$, respectively, called the ``concentration threshold'', such that
  if the problem data $b^E, b^I$ is scaled by {\tmem{any}} factor $c \geqslant
  \hat{c}$, the number of assignments/sequences that result in the optimal
  count vector $\nu^{\ast}$ is at least $1 / \varepsilon$ times greater than
  the number of {\tmem{all}} assignments that result in count vectors with
  entropy less than $(1 - \eta) G^{\ast}$ or farther than $\vartheta$ from
  $x^{\ast}$ by $\ell_1$ norm.
\end{quoteenv}

\subsubsection*{Significance}

In a problem where the only available information is embodied in the
constraints and which otherwise admits a large number of probability vectors
as solutions, the concentration phenomenon provides a powerful argument for
the {\maxent} method, which selects a particular solution, the one with
maximum entropy, in preference to all others\footnote{{\maxent} solves the
\tmtextit{inference} problem, not the \tmtextit{decision} problem. It does
{\tmem{not}} claim that the maximum entropy object is the one to use no matter
what use one has in mind.}. Likewise, the concentration results in this paper
support the maximization of generalized entropy for problems involving general
non-negative vectors. We believe that {\maxgent} can be considered to be a
compatible extension of {\maxent}. The {\tmem{compatibility}} is that any
{\maxent} problem over the reals with constraints $A^E x = b^E, A^I x
\leqslant b^I$ can be formulated as a {\maxgent} problem of the form
(\ref{eq:maxG}) with the same constraints, plus the constraint $\sum_i x_i =
1$; both problems will have the same solution $x^{\ast} \in \mathbb{R}^m$,
and the maximum entropy $H (x^{\ast})$ will equal the maximum generalized
entropy $G (x^{\ast})$. Also, if the constraints of the {\maxgent} problem
either explicitly or implicitly fix the value of $\sum_i x_i$, then the
problem can be reduced to a {\maxent} problem over the reals. The
{\tmem{extension}} consists in the fact that {\maxgent} addresses problems
involving un-normalized vectors that cannot be formulated as {\maxent}
problems, as we saw in Example \ref{ex:first}; more examples of such problems
are given in {\textsection}\ref{sec:basic}, {\textsection}\ref{sec:gconc}, and
{\textsection}\ref{sec:dconc}.

\subsubsection*{Related work}

Our term ``generalized entropy'' for $G$ is neither imaginative nor
distinctive, and there are many other generalized entropy measures. The most
general of these are Csisz{\'a}r's $f$-entropies and $f$-divergences
{\cite{Csi1996}}, and the related $\Phi$-entropies of {\cite{BLM2013}}. Any
relationship of $G$ to $\Phi$-entropies remains to be investigated. The
function $G$, in the form of the log of a multinomial coefficient with
``variable numerator'', appeared in {\cite{OSVPN2006}} and {\cite{Oik2010}}.

The problem of inferring a non-negative real vector from information in the
form of linear equalities was considered by Skilling {\cite{Skilling1989}},
where such vectors were termed ``positive additive distributions'', and by
Csizs{\'a}r, {\cite{Csi1991}}, {\cite{Csi1996}}. Both authors gave axiomatic
justifications, which do not involve probabilities, for minimizing the
I-divergence, a generalization of relative entropy to un-normalized vectors. A
further generalization is the $\alpha, \beta$ divergences of {\cite{NMF2011}}.
We discuss a connection between I-divergence and our generalized entropy in
{\textsection}\ref{sec:idiv}.

With respect to concentration, recent developments for the discrete,
normalized case were given in {\cite{entc2016}}. The continuous normalized
case, for relative entropy, is examined in {\cite{CatEIFP}} from the viewpoint
of information geometry. Countable spaces are also treated in {\cite{Gr2008}}.
But these references do not provide explicit bounds such as the ones here and
in {\cite{entc2016}}. To our knowledge, concentration for non-density vectors
has not been studied before.

The structure and some of the presentation of this paper are similar to
{\cite{entc2016}} because of the similar subject matter, entropy concentration
from a combinatorial viewpoint. Many of the results here that appear similar
to those of section III of {\cite{entc2016}} are generalizations of those
results, insofar as $G$ is a generalization of $H$. However the main theorems
here do not actually subsume corresponding theorems in {\cite{entc2016}},
because in both cases the theorems include optimizations specific to count or
frequency vectors, respectively.

\section{The generalized entropy $G$\label{sec:G}}

In this section we introduce the generalized entropy function $G$, and study
its properties, relationships with other functions, and its maximization under
linear constraints.

\begin{figure}[h]
  \resizebox{7.5cm}{!}{\includegraphics{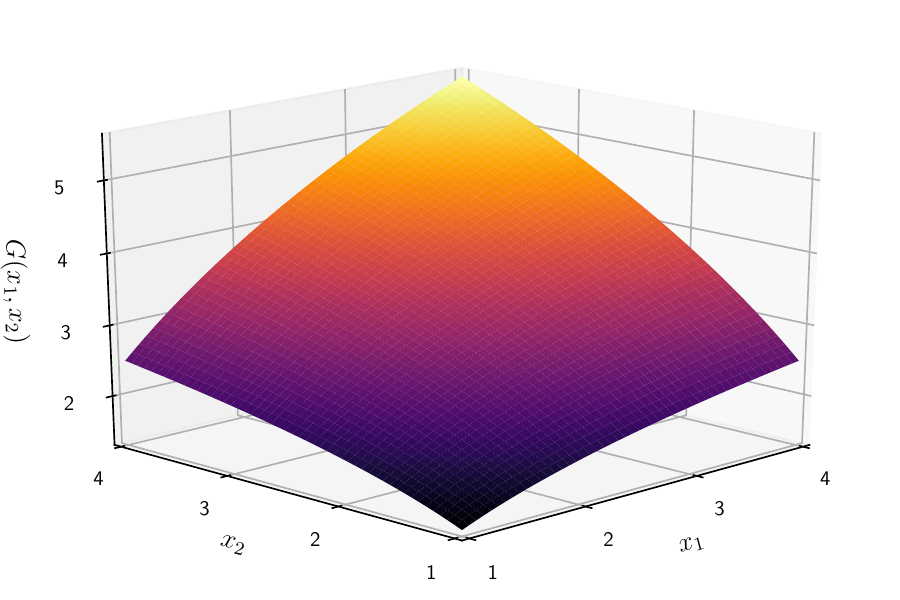}}
  \resizebox{7.5cm}{!}{\includegraphics{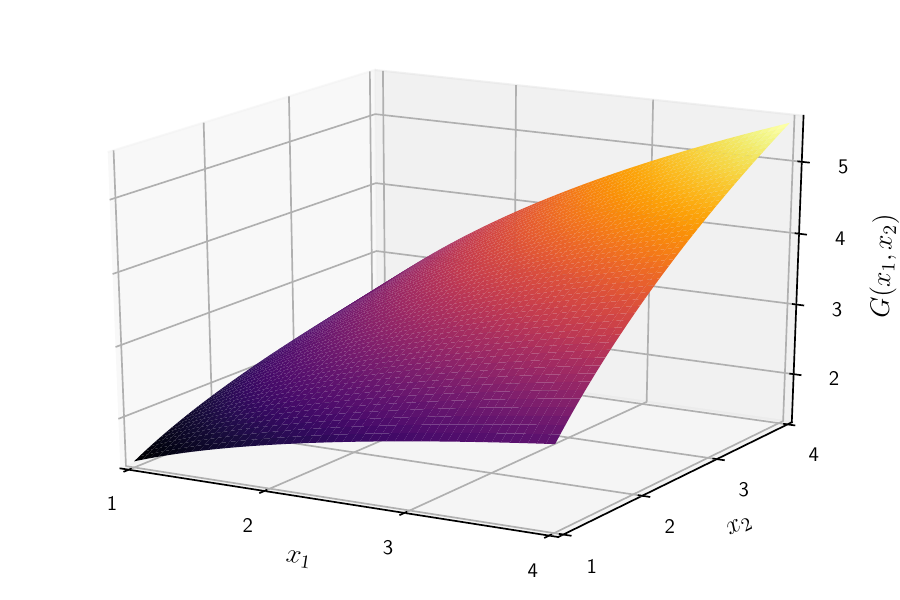}}
  \caption{\label{fig:G2}$G (x_1, x_2)$. Note that $G (x, x) = (2 \ln 2) x$,
  which destroys the strict concavity of $G$.}
\end{figure}

Given a real vector $x \geqslant 0$, its \tmtextit{generalized entropy} is
\begin{equation}
  G (x) \; \triangleq \; H (x) + \Bigl( \sum_i x_i \Bigr) \ln \Bigl( \sum_i
  x_i \Bigr) \; = \; \Bigl( \sum_i x_i \Bigr) H (\chi), \quad x \geqslant 0.
  \label{eq:G}
\end{equation}
Here $H (x)$ is the form $- \sum_i x_i \ln x_i$ extended to vectors in
$\mathbb{R}_+^m$ that are not necessarily density vectors, and $\chi$ is the
density, or normalized, or probability, vector corresponding to $x$.
(\ref{eq:G}) gives two ways to look at $G (x)$: it is the (extended) entropy
of $x$ plus the sum of $x$ times its log, or the sum of $x$ times the ordinary
entropy of the normalized $x$. If $x$ is already normalized $G (x)$ coincides
with $H (x)$. Fig. \ref{fig:G2} is a plot of $G (x)$ for $m = 2$.

\subsection{Basic properties\label{sec:Gbasic}}

We list some important properties of the function $G$:

\begin{descriptioncompact}
  \item[P1] $G (x_1, \ldots, x_m)$ is the log of the multinomial coefficient
  $\binom{x_1 + \cdots + x_m}{x_1, \ldots, x_m}$ to ``second Stirling order'':
  by using the first two terms of $\ln x! \; = \; x \ln x - x + \frac{1}{2}
  \ln x + \ln \sqrt{2 \pi} + \frac{\vartheta}{12 x}, \vartheta \in (0, 1),$we
  find that
  \[ \ln \binom{x_1 + \cdots + x_m}{x_1, \ldots, x_m} \approx G (x_1, \ldots,
     x_m) . \]
  This interpretation was given in {\cite{Oik2010}}, where it was used to
  derive ``most likely'' matrices, i.e. those with the largest number of
  realizations, from incomplete information.
  
  \item[P2] $G$ is related to the ordinary entropy (of density vectors) and
  the extended entropy (of arbitrary non-negative vectors) $H$ in the two ways
  specified in (\ref{eq:G}).
  
  \item[P3] \label{Gpropinc}Unlike the entropy of normalized vectors which is
  bounded by $\ln m$, the generalized entropy $G (x)$ increases without bound
  as the elements of $x$ become larger: for any $x, y$, if $y \geqslant x$
  then $G (y) \geqslant G (x)$. This is shown in Proposition \ref{prop:max}.
  One consequence is that if $x, y$ are close in norm, i.e. $\|x - y\|
  \leqslant \zeta$, $|G (x) - G (y) |$ cannot be bounded by an expression
  involving only $m$ and $\zeta$.
  
  \item[P4] $G (x)$ is positive, unless $x$ has just one non-0 element, in
  which case $G (x) = 0$. This follows from the second form in (\ref{eq:G}).
  
  \item[P5] \label{Gpropp}Given any p.d. $p = (p_1, \ldots, p_m)$ and any
  $n$-sequence $\sigma$ with count vector $\nu$, the probability of $\sigma$
  under $p$ can be written as
  \[ {\Pr}_p (\sigma) = e^{- (G (\nu) + nD (f\|p))} \]
  where $D (\cdot \mid \cdot)$ is the divergence, or relative entropy,
  between two probability vectors and $f = \nu / n$ is the frequency vector
  corresponding to $\nu$. By substituting $G (\nu) = nH (f)$ we obtain the
  well-known expression for the same probability in terms of the ordinary
  entropy of a frequency vector.
  
  \item[P6] Like the ordinary or the extended $H,$ $G (x_1, \ldots, x_m)$ is
  {\tmem{concave}} over the domain $x_1 > 0, \ldots, x_m > 0$, but unlike $H$,
  it is not strictly concave. See Proposition \ref{prop:Gconc} in
  {\textsection}\ref{sec:Gconc}.
  
  \item[P7] \label{Gpropmax}The maximum of $G (x_1, \ldots, x_m)$ subject just
  to the constraint $\sum_i x_i = s$ is $s \ln m$. When $s = 1$, $x$ is a
  density vector and this reduces to the maximum of $H$.
  
  \item[P8] \label{GpropH}What is the relationship between maximizing $G$ and
  maximizing the {\tmem{extended}} $H$? Consider maximizing the first form in
  (\ref{eq:G}), subject to $A^E x = b^E, A^I x \leqslant b^I$, by imposing the
  additional constraint $\sum_i x_i = s$ and treating $s$ as a parameter
  taking values in $[s_1, s_2]$. For a given $s$, there will be a unique
  maximum since $H$ is strictly concave\footnote{One may also maximize $H$
  \tmtextit{without} the constraint $\sum x_i = s$, but what would the result
  mean?}. Further, some $s = s^{\ast}$ will achieve $\max_s \max_x (s \ln s +
  H (x))$ subject to $A^E x = b^E, A^I x \leqslant b^I, \sum_i x_i = s$; this
  maximum value will equal $G (x^{\ast})$. Using the second form in
  (\ref{eq:G}) we see that there is a similar relationship between maximizing
  $G (x)$ and maximizing the function $sH (x / s)$.
  
  \item[P9] \label{Gpropsc}$G$ has a \tmtextit{scaling} (or
  \tmtextit{homogeneity}) property, which $H$ does not: for any $c > 0$ and
  any $x \in \mathbb{R}_+^m$, $G (cx) = cG (x)$. This is most easily seen
  from the second form in (\ref{eq:G}).
  
  \item[P10] \label{Gpropsc2}$G$ has a further important scaling property: if
  $x^{\ast}$ maximizes $G (x)$ under $Ax \leqslant b$, then for any $c > 0$,
  $cx^{\ast}$ maximizes $G (x)$ under $Ax \leqslant cb$. We show this in
  {\textsection}\ref{sec:scaling}, Proposition \ref{prop:x*scale}.
\end{descriptioncompact}

\subsection{Monotonicity and concavity properties\label{sec:Gconc}}

As we noted in property \ref{Gpropinc} in {\textsection}\ref{sec:Gbasic}, $G$
is an increasing function in the sense that

\begin{proposition}
  \label{prop:inc}For any $x, y$, if $y \geqslant x$ then $G (y) \geqslant G
  (x)$, and if the inequality is strict in some places, then $G (y) > G (x)$.
\end{proposition}

We will use this property in {\textsection}\ref{sec:lb}. Now we turn to
concavity.

The extended ordinary entropy $H$ is strictly concave, and in addition,
strongly concave for any modulus $\gamma \leqslant 1 / a$ when defined over
$[0, a]^m$. The generalized entropy $G$ is also concave, but neither strictly
concave, nor strongly concave for any modulus. However $- G$ is sublinear,
whereas $- H$ is not. These properties are collected in the following
proposition:

\begin{proposition}
  \label{prop:Gconc}{\tmdummy}
  
  \begin{enumeratenumeric}
    \item The function $G (x_1, .., x_m)$ is concave over $\mathbb{R}_+^m$.
    
    \item G is not \emph{strictly} concave over $\mathbb{R}_+^m$.
    
    \item $G$ is not \emph{strongly} concave over $\mathbb{R}_+^m$ for any modulus
    $\gamma > 0$.
    
    \item If the definition of $G$ is extended over all of $\mathbb{R}^m$ by
    setting $G (x) = - \infty$ if any $x_i$ is $< \; 0$, then for all $\alpha,
    \beta > 0$ and for all $x, y \in \mathbb{R}^m$, $G (\alpha x + \beta y)
    \; \geqslant \; \alpha G (x) + \beta G (y)$.
  \end{enumeratenumeric}
\end{proposition}

The last property is stronger than (implies) concavity since $\alpha, \beta$
are not required to sum to 1. The absence of strict concavity means that more
care is needed with maximization, we address this in
{\textsection}\ref{sec:max}.

\subsection{Lower bounds\label{sec:lb}}

Given a point $x$, if some other point $y$ is close to it in the distance/norm
sense, how much smaller than $G (x)$ can $G (y)$ be? We will need the answer
in {\textsection}\ref{sec:gconc}. Proposition \ref{prop:inc} implies that if
we have a hypercube centered at $x$, say $\|x - y\|_{\infty} \leqslant \zeta$,
then $G (\cdummy)$ attains its maximum at the ``upper right-hand'' corner of
the hypercube and its minimum at the ``lower left-hand'' corner. Specifically,
for any $\zeta > 0$, let $\tmmathbf{\zeta}$ denote the $m$-vector $(\zeta,
\ldots, \zeta)$, and let $x \geqslant \tmmathbf{\zeta}$. Then it can be seen
from Proposition \ref{prop:inc} that for any $y \geqslant 0$
\begin{equation}
  \|x - y\|_{\infty} \leqslant \zeta \quad \Rightarrow \quad G (x
  -\tmmathbf{\zeta}) \leqslant G (y) \leqslant G (x +\tmmathbf{\zeta}) .
  \label{eq:hcube}
\end{equation}
Using this observation we can show that

\begin{lemma}
  \label{le:close}Given $\zeta > 0$ and $x, y \in \mathbb{R}^m_+$, if $x
  >\tmmathbf{\zeta}$ and $\| y - x \|_{\infty} \leqslant \zeta$, then
  \[ G (y) \; \geqslant \; G (x) - \Bigl( \sum_i \ln \frac{\|x\|_1}{x_i}
     \Bigr) \zeta - \frac{1}{2}  \Bigl( \sum_i \frac{1}{x_i - \zeta} -
     \frac{m}{\|x\|_1 / m - \zeta} \Bigr) \zeta^2 . \]
  The coefficient of $\zeta^2$ is positive unless all $x_i$ are equal, in
  which case it becomes 0.
\end{lemma}

The lower bound above does not depend on $y$, only on $x$ and $\zeta$. The
restriction $x >\tmmathbf{\zeta}$ applies to the `reference' point $x$, not to
the `variable' $y$; see also Remark \ref{rem:xs1}. Lastly, since $\| x - y
\|_1 \leqslant \zeta \Rightarrow \| x - y \|_{\infty} \leqslant \zeta$, the
lemma holds also when the $\ell_{\infty}$ norm is replaced by the $\ell_1$
norm.

We will use Lemma \ref{le:close} in {\textsection}\ref{sec:lbNnu*} to bound
how far from the maximum $G (x^{\ast})$ the value $G (x)$ can be if $x$ is
close to $x^{\ast}$. We also comment there, Remark \ref{rem:boundcomp}, on how
the above bound compares to bounds obtainable from the relationship between
(ordinary) entropy difference and $\ell_1$ norm.

\subsection{Maximization\label{sec:max}}

Let $\mathcal{C} (0)$ denote the subset of $\mathbb{R}^m$ defined by
the constraints in (\ref{eq:maxG})\footnote{The reason for the ``0''
will be seen in {\textsection}\ref{sec:constr}, where we discuss
tolerances on constraints.}.  Here we point out that despite the fact
that $G$ is not a strictly concave function (recall Proposition
\ref{prop:Gconc}, part 2), the point $x^{\ast}$ solving
(\ref{eq:maxG}) is the {\tmem{unique}} optimal solution of our maximization
problem, and occupies a special location in $\mathcal{C}(0)$:

\begin{proposition}
  \label{prop:max}{\tmdummy}
  \begin{enumerate}
  \item The point $x^{\ast}$ is the unique optimal solution of problem
    (\ref{eq:maxG}).
    \item The set $\mathcal{C} (0)$ does not contain any $x$ s.t. $x
    \geqslant x^{\ast}$ with at least one strict inequality.
  \end{enumerate}
\end{proposition}

\begin{figure}[h]
  \centering
  \resizebox{5cm}{!}{\includegraphics{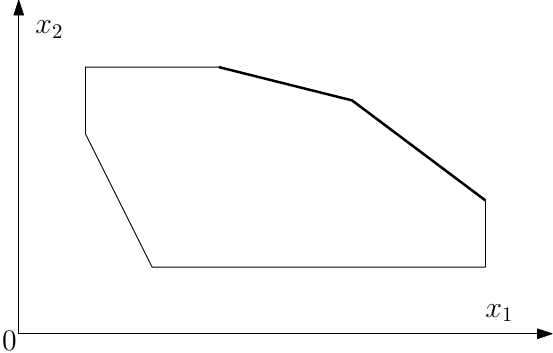}}
  \caption{\label{fig:dom}\small A 2-dimensional polytope $\mathcal{C} (0)$. By
    Proposition \ref{prop:max}, $x^{\ast}$ can lie only on the heavy black
    line.}
\end{figure}

Figure \ref{fig:dom} illustrates the first statement of the proposition.

Finally we look at the form of the solution $x^{\ast}$ in terms of Lagrange
multipliers. The Lagrangean for problem (\ref{eq:maxG}) is
\begin{equation}
  L (x, \lambda^E, \lambda^I) = G (x) - \lambda^E \cdot (A^E x - b^E) -
  \lambda^I \cdot (A^I x - b^I), \label{eq:L}
\end{equation}
where $\lambda^E, \lambda^I$ are the vectors of the Lagrange multipliers
corresponding to the equality and inequality constraints. The solution
$x^{\ast}$ will satisfy some of the inequality constraints with equality (and
these are called \tmtextit{binding} or \tmtextit{active} at $x^{\ast}$), and
some with strict inequality. It is known that multipliers $\lambda^I_j$
corresponding to inequalities non-binding at $x^{\ast}$ will be 0, while the
rest of them will be $\geqslant \; 0$ (see, e.g., {\cite{HUL1996}}, Ch. VII,
{\textsection}2.4). Thus, denoting the sub-vector of $\lambda^I$ corresponding
to binding inequalities by $\lambda^{\tmop{BI}}$ and the corresponding
sub-matrix of $A^I$ by $A^{\tmop{BI}}$, it follows from (\ref{eq:L}) that
$x^{\ast}$ can be written as
\begin{equation}
  x^{\ast}_j = (x^{\ast}_1 + \cdots + x^{\ast}_m) e^{- (\lambda^E \cdot A^E_{.
  j} + \lambda^{\tmop{BI}} \cdot A^{\tmop{BI}}_{. j})} . \label{eq:x*j}
\end{equation}
This expression determines the elements of the density vector $\chi^{\ast} =
x^{\ast} / \sum_i x^{\ast}_i$ in terms of the multipliers, but it does not
determine the vector $x^{\ast}$ itself.

\begin{remark}
  \label{rem:zeros}It is clear that the form (\ref{eq:x*j}) cannot express any
  elements of $x^{\ast}$ that are 0, if the multipliers $\lambda$ are to be
  finite. To avoid introducing special cases in the sequel to handle the
  zeros, we will assume as a convenience that any elements of the solution to
  problem (\ref{eq:maxG}) that are forced to be exactly 0 by the constraints
  are eliminated from consideration either before or after the solution is
  found. We have already alluded to this after (\ref{eq:s1s2}). Thus, whenever
  we speak of $x^{\ast}$ in what follows we will assume that all of its
  elements are positive. See Example \ref{ex:vt} in
  {\textsection}\ref{sec:dconc}. A more detailed discussion of the issue of 0s
  is in {\cite{entc2016}}, {\textsection}II.A.
\end{remark}

\begin{example}
  \label{ex:first+}Returning to Example \ref{ex:first}, it is possible to
  maximize $G$ analytically under the given constraints. Introducing real
  variables $x_1, x_2, x_3$ corresponding to $\nu_r, \nu_g, \nu_b$ and letting
  the constraints be $x_1 + x_2 = a$ and $x_2 + x_3 \leqslant b$, the solution
  turns out to be
  \[ x^{\ast}_1 = s^{\ast} - b, \quad x^{\ast}_2 = a + b - s^{\ast}, \quad
     x^{\ast}_3 = s^{\ast} - a, \qquad s^{\ast} = \frac{a + b + \sqrt{a^2 +
     b^2}}{2} . \]
  Further, the bounds $s_1, s_2$ of \ (\ref{eq:s1s2}) on the possible sums are
  $s_1 = a$ and $s_2 = a + b$. We see that the {\maxgent} solution to the
  problem is never trivial, in the sense that for all $a, b$, we have $s_1 <
  s^{\ast} < s_2$; when $a = b$ we have $s^{\ast} = \frac{1}{2}  \left( 1 +
  \frac{\sqrt{13}}{5} \right) s_2 \approx 0.861 s_2$. With $a = 4, b = 6$ we
  find $s^{\ast} = 8.61$ and $x^{\ast} = (2.61, 1.39, 4.61)$; compare with
  Table \ref{tab:ass}. 
\end{example}

\subsection{A connection with I-divergence\label{sec:idiv}}

For density vectors, the relationship between ordinary entropy $H (x)$ and
divergence $D (x\|y)$ is well known: with uniform $y$, $D (x\|y)$ reduces to
$H (x)$ to within a constant, and its minimization is equivalent to the
maximization of $H (x)$. Here we look at whether $G (x)$ has any analogous
properties.

First, if in $D (x \| y \nobracket)$ we take $y$ to have all of its elements
equal to $\sum_i x_i$, we obtain $- G (x)$. However, this is merely a
{\tmem{formal}} relationship\footnote{This is pointed out in {\cite{BV}}, Ch.
3, Example 3.19.}. For example, minimizing $D (x \| y \nobracket)$ with
respect to $x$ when $y = \left( \sum_i x_i, \ldots, \sum_i x_i \right)$ cannot
be given the same interpretation as minimizing $D (x \| y \nobracket)$ with
respect to $x$ given a {\tmem{fixed}} `prior' $y$. So even if $x, y$ summed to
1, neither the axiomatic nor the concentration justifications for
cross-entropy minimization would apply.

Second, the concentration properties we establish in
{\textsection}\ref{sec:gconc} and {\textsection}\ref{sec:dconc} support the
maximization of $G (x)$ as a method of inference of non-negative vectors from
limited information. Another method for doing this, suggested in
{\cite{Skilling1989}}, {\cite{Csi1996}}, is based on minimizing the
{\tmem{I-divergence}} (information divergence) between non-negative vectors
\begin{equation}
  \mathcal{D} (u \| v \nobracket) \; \triangleq \; \sum_i u_i \ln
  \frac{u_i}{v_i} - \sum_i u_i + \sum_i v_i \label{eq:idiv}, \qquad u, v \in
  \mathbb{R}^m_+ .
\end{equation}
This reduces to $D (u\|v)$ when $u, v$ sum to 1. The inference problem is
``problem (iii)'' in {\cite{Csi1996}}: infer a non-negative function $p (z)$,
not necessarily summing or integrating to 1, given that (a) it belongs to a
certain feasible set $\mathcal{F}$ of functions defined by linear
{\tmem{equality}} constraints, and (b) a default model $q (z)$\footnote{The
sense of `default' is that if $q$ is in $\mathcal{F}$, then, in the absence of
any constraints, the method should infer $p^{\ast} = q$.}. It is shown that
the solution of this problem is the $p^{\ast} \in \mathcal{F}$ that minimizes
the I-divergence $\mathcal{D} (p \| q \nobracket)$. (Recently, minimization of
I-divergence and generalizations to ``$\alpha, \beta$ divergences'' has found
many applications in the area known as ``non-negative matrix factorization'',
see {\cite{NMF2011}}.)

There is a relationship between minimizing I-divergence and maximizing
generalized entropy:

\begin{proposition}
  \label{prop:idiv}Let $(A^E, b^E), (A^I, b^I)$ be linear equality and
  inequality constraints on a vector in $\mathbb{R}_+^m$, and let $x^{\ast}$
  be the solution of the {\maxgent} problem with these constraints on $x$.
  Given a prior $v \in \mathbb{R}^m_+$, let $u^{\ast} (v)$ be the solution to
  the minimum I-divergence problem with the same constraints on $u$. Then
  there is a prior $\tilde{v}$ which makes the two solutions coincide, i.e.
  $u^{\ast} (\tilde{v}) = x^{\ast}$. That prior is $\tilde{v} = (s^{\ast},
  \ldots, s^{\ast})$.
\end{proposition}

This follows from the fact that the minimum I-divergence solution to a problem
with prior $v$ and constraints $A^E u = b^E$ and $A^I u \leqslant b^I$ on $u$
is
\begin{equation}
  u^{\ast}_j = v_j e^{- (\lambda^E \cdot A^E_{. j} + \lambda^{\tmop{BI}} \cdot
  A^{\tmop{BI}}_{. j})} . \label{eq:idiv2}
\end{equation}
If we set $v_j = s^{\ast}$, it can be seen from expression (\ref{eq:x*j}) that
$u^{\ast}_j = x^{\ast}_j$ satisfies (\ref{eq:idiv2}).

Inference by minimizing I-divergence under equality constraints has an
axiomatic basis, but as pointed out in {\textsection}3 and {\textsection}7 of
{\cite{Csi1996}}, the combinatorial, concentration rationale that we are
advocating here {\tmem{does not seem to apply}} to it. Proposition
\ref{prop:idiv} shows that the adoption of a {\tmem{particular prior}}
furnishes this rationale, except that this prior cannot be properly viewed as
{\tmem{independent}} of the solution (posterior) $u^{\ast}$. This dependence
may shed some light on the difficulty of finding the concentration rationale
in general. [As an illustration, Example \ref{ex:first+} can be solved by
I-divergence minimization assuming a constant prior $v = (\alpha, \alpha,
\alpha)$. An analytical solution $u^{\ast}$ is possible, and it has the same
form as the {\maxgent} solution, but it is a function of $\alpha \in (0,
\infty)$; the question then becomes what value to adopt for $\alpha$.]

\section{Constraints, scaling, sensitivity, and the optimal count
vector\label{sec:basic}}

In {\textsection}\ref{sec:constr} we discuss the necessity of introducing
{\tmem{tolerances}} into the constraints defining the {\maxgent} problem, and
in {\textsection}\ref{sec:tolopt} the effect of these tolerances on the
maximization of $G$. In {\textsection}\ref{sec:scaling} we turn to the
{\tmem{scaling}} of the problem, i.e. multiplying the data vector $b$ by some
$c > 0$, and the important properties of this scaling. Lastly, in
{\textsection}\ref{sec:nu*} we discuss the {\tmem{optimal}}, or {\maxgent}
count vector $\nu^{\ast}$, constructed from the real vector $x^{\ast}$ solving
problem (\ref{eq:maxG}).

\subsection{Constraints with tolerances\label{sec:constr}}

We pointed out the necessity of introducing {\tmem{tolerances}} into linear
constraints when establishing concentration of ordinary entropy in
{\cite{entc2016}}. The constraints involved real coefficients, and the
solutions had to be rational (frequency) vectors with a particular
denominator. Here the solutions need to be integral (count) vectors, but the
equality constraints may not have any integral solution; e.g. $x_1 - x_2 = 1,
x_1 + x_2 = 4$ are satisfied only for $(x_1, x_2) = (2.5, 1.5)$, and likewise
with inequalities, e.g. $1.3 \leqslant x_1 \leqslant 1.99$. We therefore
define the set of real $m$-vectors $x$ that satisfy the constraints in
(\ref{eq:maxG}) with a relative accuracy or tolerance $\delta \geqslant 0$:
\begin{equation}
  \nobracket \mathcal{C} (\delta) \triangleq \{ x \in \mathbb{R}^m : b^E -
  \delta | \beta^E | \leqslant A^E x \nobracket \leqslant b^E + \delta |
  \beta^E |, A^I x \leqslant b^I + \delta | \beta^I | \}, \label{eq:Cdelta}
\end{equation}
where $\beta^E, \beta^I$ are identical to $b^E, b^I$, except that any elements
that are 0 are replaced by appropriate small positive constants. The
tolerances are only on the {\tmem{values}} $b$ of the constraints, not on
their structure $A$. Recall that the generalized entropy is maximized over
$\mathcal{C} (0)$, which we have assumed to be non-empty, problem
(\ref{eq:maxG}).

There are three main points concerning the introduction of $\delta$. First,
the existence of integral solutions, which is elaborated in Proposition
\ref{prop:theta_inf} below. Second, and related to the first, $\delta$ ensures
that the concentration statement {\tmem{GC}} in {\textsection}\ref{sec:intro}
holds {\tmem{for all}} scalings of the problem larger than a threshold
$\hat{c}$. This is analogous to having concentration for frequency (rational)
vectors hold for all denominators $n$ larger than some $N$, as in
{\cite{entc2016}}. Third, $\delta$ has an effect on the maximization of $G$;
this the subject of {\textsection}\ref{sec:tolopt}.

Proposition \ref{prop:theta_inf} below gives the fundamental facts about the
existence of count vectors in $\mathcal{C} (\delta)$. Given an $x$ in
$\mathcal{C} (0)$, any other vector $y$ close enough to it is in $\mathcal{C}
(\delta)$, and, if $\delta$ is not too small, the {\tmem{count}} vector
obtained by rounding $x$ element-wise is in $\mathcal{C} (\delta)$; in other
words, for every real vector in $\mathcal{C} (0)$ there is an integral vector
in $\mathcal{C} (\delta)$. The ``close enough'' and the ``not too small''
depend on a number $\vartheta_{\infty}$:

\begin{proposition}
  \label{prop:theta_inf}With $\beta^E, \beta^I$ as in (\ref{eq:Cdelta}),
  define
  \[ \vartheta_{\infty} \; \triangleq \; \min (| \beta^E |_{\min} /
     \interleave A^E \interleave_{\infty}, | \beta^I |_{\min} / \interleave
     A^I \interleave_{\infty}), \]
  or $\infty$ if there are no constraints\footnote{Recall that the infinity
  norm $\interleave \cdot \interleave_{\infty}$ of a matrix is the maximum of
  the $\ell_1$ norms of the rows.}. Then if $x$ is any point in $\mathcal{C}
  (0)$,
  \begin{enumeratenumeric}
    \item Given any $\delta > 0$, any $y \in \mathbb{R}_+^m$ such that $\|y -
    x\|_{\infty} \; \leqslant \delta \; \vartheta_{\infty}$ is in $\mathcal{C}
    (\delta)$.
    
    \item In particular, if $\delta \geqslant 1 / (2 \vartheta_{\infty})$, the
    integral/count vector $[x]$ is in $\mathcal{C} (\delta)$.
  \end{enumeratenumeric}
\end{proposition}

As we add constraints to a problem, $\vartheta_{\infty}$ can only decrease, or
at best stay the same. This proposition is used in
{\textsection}\ref{sec:scale}, eq. (\ref{eq:cref3}), and in
{\textsection}\ref{sec:cdist}, after (\ref{eq:nlb3}).

\begin{example}
  \label{ex:imp}Fig. \ref{fig:imp} shows a network consisting of 6 nodes and 6
  links. The links are subject to a certain impairment $x$ and $x_i$ is the
  quantity associated with link $i$. The impairment is {\tmem{additive}}, e.g.
  its value over the path $A B$ consisting of links $4, 1, 6$ is $x_4 + x_1 +
  x_6$.
  
  \begin{figure}[h]
    \centering
    \raisebox{-0.386058504968723\height}{\includegraphics{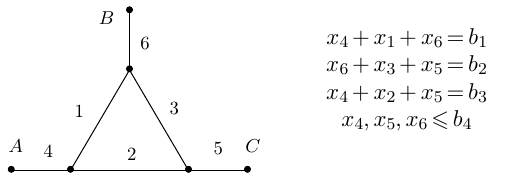}}
    \caption{\label{fig:imp}\small Data $b$ on the impairment $x$ in a 6-node,
      6-link network.}
  \end{figure} Suppose that $x$ is measured over the 3 paths $A B, B C, C A$,
  and it is also known that the access links 4, 5, 6 contribute no more than a
  certain amount, as shown in Fig. \ref{fig:imp}. The structure matrices $A^E,
  A^I$ and data vectors $b^E, b^I$ then are
  \begin{gather*}\small
    A^E = \left[
      \begin{array}{cccccc}
       1 & 0 & 0 & 1 & 0 & 1\\
       0 & 0 & 1 & 0 & 1 & 1\\
       0 & 1 & 0 & 1 & 1 & 0
    \end{array}\right], \quad
    b^E = \left[
      \begin{array}{c}
       b_1\\
       b_2\\
       b_3
    \end{array}\right], \quad
    A^I = \left[
      \begin{array}{cccccc}
       0 & 0 & 0 & 1 & 0 & 0\\
       0 & 0 & 0 & 0 & 1 & 0\\
       0 & 0 & 0 & 0 & 0 & 1
    \end{array}\right], \quad
    b^I = \left[
      \begin{array}{c}
       b_4\\
       b_4\\
       b_4
    \end{array}\right] .
  \end{gather*}
  The problem is to infer the impairment vector $x$ from the measurement
  vector $b$. Clearly, the values of the $b_i$ depend on the chosen units and
  can change under various conditions, whereas the elements of $A^E, A^I$ are
  constants defining the structure of the network, and independent of any
  units.
  
  Suppose we take $(b_1, \ldots, b_4) = (10.5, 18.3, 8.7, 4)$. Then with
  $\beta^E, \beta^I = b^E, b^I$ we have in Proposition \ref{prop:theta_inf} $|
  b^E |_{\min} / \interleave A^E \interleave_{\infty} = 8.7 / 3$ and $| b^I
  |_{\min} / \interleave A^I \interleave_{\infty} = 4 / 1$, so
  $\vartheta_{\infty} = 2.9$. The vector $x = (6.591, 5.326, 13.26, 1.120,
  2.253, 2.789)$ satisfies the constraints exactly. The rounded vector $[x] =
  (7, 5, 13, 1, 2, 3)$ is in the set $\mathcal{C} (\delta)$ defined by
  (\ref{eq:Cdelta}) for any $\delta \geqslant 0.172$.
\end{example}

\subsection{Effect of tolerances on the optimality of
$x^{\ast}$\label{sec:tolopt}}

With the constraints $x_1 - x_2 = 1, x_1 + x_2 = 4,$and $x_1, x_2 \geqslant
0$, $\mathcal{C} (0)$ is a 0-dimensional polytope in $\mathbb{R}^2$, the
point $(2.5, 1.5)$. However, introducing the tolerance \ $\delta = 0.05$ turns
the equalities into inequalities and $\mathcal{C} (0.05) = \{0.95 \leqslant
x_1 - x_2 \leqslant 1.05, 3.8 \leqslant x_1 + x_2 \leqslant 4.2\}$ becomes
2-dimensional. Apart from the change in dimension, $\mathcal{C} (0.05)$ also
contains the point $(2.55, 1.55)$ at which $G$ assumes a value
{\tmem{greater}} than $G^{\ast} = G (2.5, 1.5)$, its maximum over $\mathcal{C}
(0)$. This must be taken into account, since concentration refers to the
vectors in $\mathcal{C} (\delta)$, not those in $\mathcal{C} (0)$. The
following lemma shows that the amount by which the value of $G$ can exceed
$G^{\ast}$ due to the widening of the domain $\mathcal{C} (0)$ to $\mathcal{C}
(\delta)$ is bounded by a {\tmem{linear}} function of $\delta$; it generalizes
Prop. II.2 of {\cite{entc2016}} for the ordinary entropy $H$:

\begin{lemma}
  \label{le:far}Let $(\lambda^E, \lambda^{\tmop{BI}})$ be the vector of
  Lagrange multipliers in (\ref{eq:x*j}) corresponding to the solution
  $G^{\ast} = G^{\ast} (0), x^{\ast} = x^{\ast} (0)$ of the maximization
  problem (\ref{eq:G}). Define
  \[ \Lambda^{\ast} \; \triangleq \; | \lambda^E | \cdot |\beta^E | +
     \lambda^{\tmop{BI}} \cdot |\beta^{\tmop{BI}} |, \qquad \Lambda^{\ast}
     \geqslant G^{\ast} . \]
  Then with $\delta \geqslant 0$, for any $\nu \in \mathcal{C} (\delta)$
  \[ G (\nu) \; \leqslant \; G^{\ast} + \Lambda^{\ast} \delta - nD (f\|
     \chi^{\ast}), \]
  where $n = \sum_i \nu_i$, $f$ is the frequency vector corresponding to
  $\nu$, and $\chi^{\ast}$ is the density vector corresponding to $x^{\ast}$.
\end{lemma}

The upper bound on $G (\nu)$ is at least $(1 + \delta) G^{\ast} - nD (f\|
\chi^{\ast})$. When $\delta = 0$ the lemma says simply that $G^{\ast}$ is the
maximum of $G$ over $\mathcal{C} (0)$. The $D (\cdot \| \cdot \nobracket)$
term is positive, and equals 0 iff $\nu = \alpha x^{\ast}$ for some $\alpha >
0$\footnote{The only way the density vectors can be equal is if the
un-normalized vectors are proportional.}. Leaving aside that this is possible
only for special $x^{\ast}$ and $\alpha$, Lemma \ref{le:far} says that if the
resulting $\nu$ is in $\mathcal{C} (\delta)$, then $G (\nu) = \alpha G^{\ast}
\leqslant G^{\ast} + \Lambda^{\ast} \delta$, i.e. the allowable $\alpha$ is
limited by $\delta$. Also, if we have even one equality constraint, $\delta$
limits the size of the allowable $\alpha$ even further.

\subsection{Scaling of the data and bounds on the allowable
sums\label{sec:scaling}}

We establish a fundamental property, \ref{Gpropsc2} in
{\textsection}\ref{sec:Gbasic}, of maximizing the generalized entropy $G$: if
the problem data $b$ is scaled by the factor $c > 0$, all aspects of the
solution scale by the same factor.

\begin{proposition}
  \label{prop:x*scale}Suppose that the relaxed count vector $x^{\ast}$
  maximizes $G (x)$ under the linear constraints $A^E x = b^E, A^I x \leqslant
  b^I$, which also imply that $\sum_i x_i$ is between the bounds $s_1, s_2$.
  Let $c > 0$ be any constant. Then the vector $cx^{\ast}$ maximizes $G (x)$
  under the scaled constraints $A^E x = cb^E, A^I x \leqslant cb^I$, the
  maximum value of $G$ is $cG (x^{\ast})$, and the new bounds on $\sum_i x_i$
  are $cs_1, cs_2$.
\end{proposition}

How do $s_1$ and $s_2$, defined in (\ref{eq:s1s2}), depend on the structure
matrices $A^E, A^I$ and the data $b^E, b^I$? In general, the problem of
bounding $s_1$ or $s_2$ doesn't have a simple answer: by scaling the
variables, {\tmem{any}} linear program whose objective function is a positive
linear combination of the variables can be converted to one where the
objective function is simply the sum of the variables. But in some special
cases we can derive simple bounds on $s_1$ and $s_2$:
\begin{proposition}
  \label{prop:s1s2bounds}Bounds on the sums $s_1$ and $s_2$.
  \begin{enumerate}
    \item If there are some equality constraints, then $s_1 \geqslant \|b^E
    \|_1 / \interleave (A^E)^T \interleave_{\infty}$. (This bound can only
    increase if there are also inequalities.)
    \item Suppose all of $A^E, A^I, b^E, b^I$ are $\geqslant 0$, and each
    $x_i$ occurs in at least one constraint. Then $s_2 \leqslant \sum_i b^E_i
    / \alpha^E_i + \sum_i b^I_i / \alpha^I_i$, where $\alpha^E_i$,
    $\alpha^I_i$, is the smallest non-zero element of row $i$ of $A^E$,
    respectively $A^I$, if that element is $< 1$, and 1 otherwise.
  \end{enumerate}
\end{proposition}
Recall from {\textsection}\ref{sec:intro} that ``each $x_i$ occurs in at least
one constraint'' is a necessary condition for the problem to be bounded. The
proposition applies to Example \ref{ex:imp}: we find that $s_1 \geqslant (b_1
+ b_2 + b_3) / 2$ and $s_2 \leqslant b_1 + b_2 + b_3 + b_4$.

\subsection{The optimal count vector $\nu^{\ast}$\label{sec:nu*}}

Given the relaxed optimal count vector $x^{\ast}$, we construct from it a
{\tmem{count}} vector $\nu^{\ast}$ which is a reasonable approximation to the
integral vector that solves problem (\ref{eq:maxG}), in the sense that (a) its
sum is close to that of $x^{\ast}$, and (b) its distance from $x^{\ast}$ is
small in $\ell_1$ norm. These properties will be needed in
{\textsection}\ref{sec:gconc} and {\textsection}\ref{sec:dconc}. We will
require $\nu^{\ast}$ to sum to $n^{\ast}$, where
\begin{equation}
  n^{\ast} \triangleq \lceil s^{\ast} \rceil, \quad s^{\ast} \triangleq \sum_i
  x^{\ast}_i . \label{eq:n*s*}
\end{equation}
For any $x \geqslant 0$, let $[x]$ be the vector obtained by rounding each of
the elements of $x$ up or down to the nearest integer.  $\nu^{\ast}$ is
obtained from $x^{\ast}$ by a process of rounding and adjusting:
\vspace*{-2ex}
\begin{definition}[{\cite{entc2016}}, Defn. III.1]
  \label{def:nu*}Given $x^{\ast}$, form the density vector $\chi^{\ast} =
  x^{\ast} / s^{\ast}$ and set $\tilde{\nu} = [n^{\ast} \chi^{\ast}]$.
  Construct $\nu^{\ast}$ by adjusting $\tilde{\nu}$ as follows. Let $d = \sum_i
  \tilde{\nu}_i - n^{\ast} \in \mathbb{Z}$. If $d = 0$, set $\nu^{\ast} =
  \tilde{\nu}$. Otherwise, if $d < 0$, add 1 to $| d |$ elements of
  $\tilde{\nu}$ that were rounded down, and if $d > 0$, subtract 1 from $| d |$
  elements that were rounded up.  The resulting vector is $\nu^{\ast}$.
\end{definition}
We will refer to $\nu^{\ast}$ as ``the optimal count vector'' or ``the
{\maxgent} count vector'' (even though it is not unique). It sums to
$n^{\ast}$, and does not differ too much from $x^{\ast}$ in norm:
\begin{proposition}
  \label{prop:nu*}The optimal count vector $\nu^{\ast}$ of Definition
  \ref{def:nu*} is such that
  \[ \sum_{1 \leqslant i \leqslant m} \nu^{\ast}_i = n^{\ast}, \quad \|
     \nu^{\ast} - x^{\ast} \|_1 \leqslant \frac{3 m}{4} + 1, \quad \|
     \nu^{\ast} - x^{\ast} \|_{\infty} \leqslant 1, \quad \| f^{\ast} -
     \chi^{\ast} \|_1 \leqslant \frac{3 m}{4 n^{\ast}} . \]
\end{proposition}
There are other approximations to the integral solution of problem
(\ref{eq:maxG}); for example, simply $[x^{\ast}]$ achieves smaller norms than
$\nu^{\ast}$: $\| [x^{\ast}] - x^{\ast} \|_1 \leqslant m / 2$, $\| [x^{\ast}]
- x^{\ast} \|_{\infty} \leqslant 1 / 2$. ($[x^{\ast}]$ is the point of
$\mathbb{N}^m$ that minimizes the Euclidean distance $\| \nu - x^{\ast} \|_2$
from $x^{\ast}$.) But $[x^{\ast}]$ does not have the required sum $n^{\ast}$.

Another, more sophisticated definition for $\nu^{\ast}$, would use the
solution of the integer linear program $\min_{\nu \in \mathbb{N}^m} \sum_{i =
1}^m | \nu_i - x^{\ast}_i |$ subject to $\sum_{i = 1}^m \nu_i = n^{\ast}$.
[This is a linear program because $\min_z  \sum_i |z_i - c_i |$ is equivalent
to $\min_{a, z}  \sum_i a_i$ subject to $a_i \ge z_i - c_i, a_i \ge -(z_i -
c_i)$.] A $\nu^{\ast}$ better than that of Definition \ref{def:nu*} would
improve the bound in (\ref{eq:Gnu*lb}) below\footnote{No integral vector can
achieve $\ell_1$ norm smaller than $\| x^{\ast} - [x^{\ast}] \|_1$; this
solution to the linear program ignores the constraint, and minimizes each term
of the objective function individually.}.

\section{Concentration with respect to entropy difference\label{sec:gconc}}

It is not clear that concentration should occur at all in a situation like the
one of Example \ref{ex:first}. The fact that $G$ has a global maximum
$G^{\ast}$ over $\mathcal{C} (0)$ is not enough. In this section we
demonstrate that concentration around $G^{\ast}$ does indeed occur, in the
sense of the statement {\tmem{GC}} of {\textsection}\ref{sec:intro},
pertaining to entropies $\eta$-far from $G^{\ast}$. This is done in two
stages, by Theorem \ref{th:Gdiff} in {\textsection}\ref{sec:ubNB} and Theorem
\ref{th:cGdiff} in {\textsection}\ref{sec:scale}.

Consider the count vectors that sum to $n$ and satisfy the constraints. We
divide them into two sets, $\mathcal{A}_n, \mathcal{B}_n$, according to the
{\tmem{deviation of their generalized entropy}} from $G^{\ast}$: given
$\delta, \eta > 0$,
\begin{equation}
  \begin{array}{lll}
    \mathcal{A}_n (\delta, \eta) & \triangleq & \{ \nu \in N_n \cap
    \mathcal{C}(\delta), G (\nu) \geqslant (1 - \eta) G^{\ast} \},\\
    \mathcal{B}_n (\delta, \eta) & \triangleq & \{ \nu \in N_n \cap
    \mathcal{C}(\delta), G (\nu) < (1 - \eta) G^{\ast} \} .
  \end{array} \label{eq:AnBneta}
\end{equation}
Irrespective of the values of $\delta$ and $\eta$, $\mathcal{A}_n (\delta,
\eta) \uplus \mathcal{B}_n (\delta, \eta) = N_n \cap \mathcal{C} (\delta)$.
Now we discuss the possible range of $n$.

We have assumed that the problem constraints $A^E x = b^E, A^I x \leqslant
b^I$ imply that $s_1 \leqslant \sum_i x_i \leqslant s_2$, where the bounds
$s_1, s_2$ on the sum of $x$ are found by solving the linear programs
(\ref{eq:s1s2}). So any integral vector that satisfies the constraints
{\tmem{exactly}}, i.e. is in $\mathcal{C} (0)$, must have a sum $n$ between
$n_1 = \lceil s_1 \rceil$ and $n_2 = \lfloor s_2 \rfloor$. We will use a
slight modification of this definition
\begin{equation}
  n_1 \triangleq \lceil s_1 \rceil, \quad n_2 \triangleq \lceil s_2 \rceil .
  \label{eq:n1n2}
\end{equation}
With $n^{\ast}$ defined by (\ref{eq:n*s*}), we have $n_1 \leqslant n^{\ast}
\leqslant n_2$. We may assume without loss of generality that $n_1 \leqslant
n_2 + 1$; otherwise all count vectors sum to a known $n$, and we reduce to the
case of frequency vectors which was studied in {\cite{entc2016}}.

\begin{remark}
  There is a certain degree of arbitrariness (or flexibility) in the
  definitions of $n_1, n_2$. Setting $n_1 = \lceil s_1 \rceil, n_2 = \lfloor
  s_2 \rfloor$ says that the allowable {\tmem{sums}} are those of count
  vectors which belong to $\mathcal{C} (0)$; it does not say that the only
  allowable {\tmem{vectors}} are those in $\mathcal{C} (0)$. Now it could be
  argued that after introducing the tolerance $\delta$, the numbers $n_1, n_2$
  should be allowed to become functions of $\delta$. However, this would
  introduce significant extra complexity. Our definition makes concessions to
  simplicity by restricting somewhat the allowable sums, and by slightly
  adjusting the value of $n_2$ to handle the `boundary' case $\lfloor s_2
  \rfloor < s^{\ast} \leqslant s_2$ more easily.
\end{remark}

Having defined the range of allowable sums $n$ as $n_1 \leqslant n \leqslant
n_2$, we will use the (disjoint) unions of the sets (\ref{eq:AnBneta}) over $n
\in \{ n_1, \ldots, n_2 \}$
\begin{equation}
  \begin{array}{lll}
    \mathcal{A}_{n_1 : n_2} (\delta, \eta) & \triangleq & \left\{ \nu \middle|
    \; \sum_i \nu_i = n, n_1 \leqslant n \leqslant n_2, \nu \in \mathcal{C}
    (\delta), G (\nu) \geqslant (1 - \eta) G^{\ast} \right\},\\
    \mathcal{B}_{n_1 : n_2} (\delta, \eta) & \triangleq & \left\{ \nu \middle|
    \; \sum_i \nu_i = n, n_1 \leqslant n \leqslant n_2, \nu \in \mathcal{C}
    (\delta), G (\nu) < (1 - \eta) G^{\ast} \right\} .
  \end{array} \label{eq:ABn1n2}
\end{equation}
Irrespective of $\delta$ and $\eta$ we have
\begin{equation}
  \mathcal{A}_{n_1 : n_2} (\delta, \eta) \; \uplus \; \mathcal{B}_{n_1 : n_2}
  (\delta, \eta) \; = \; N_{n_1 : n_2} \cap \mathcal{C} (\delta) .
  \label{eq:Nn1n2}
\end{equation}
We note the following relationship among the numbers of realizations of the
optimal count vector $\nu^{\ast}$ and those of the sets $\mathcal{A}_{n_1 :
n_2} (\delta, \eta)$ and $\mathcal{B}_{n_1 : n_2} (\delta, \eta)$: if
$\nu^{\ast} \in \mathcal{A}_{n_1 : n_2} (\delta, \eta)$, then
\begin{equation}
  \begin{array}{rrrrrrrrr}
    \displaystyle\frac{\# \nu^{\ast}}{\#\mathcal{B}_{n_1 : n_2} (\delta, \eta)} &
    \geqslant  &
    \displaystyle \frac{1}{\varepsilon} &
    \displaystyle \Rightarrow &
    \displaystyle \frac{\# \mathcal{A}_{n_1 : n_2} (\delta, \eta) +\# \;
      \mathcal{B}_{n_1 : n_2} (\delta, \eta)}{\# \mathcal{A}_{n_1 : n_2}
      (\delta, \eta)} &
    \displaystyle \leqslant & 1 + \varepsilon &
    \Rightarrow & \\
    &  &  &  &
    \displaystyle \; \frac{\# \mathcal{A}_{n_1 : n_2} (\delta, \eta)}{\# (N_{n_1
        : n_2} \cap \mathcal{C} (\delta))} & \geqslant &
    \displaystyle \frac{1}{1 + \varepsilon} & > &
    \displaystyle 1 - \varepsilon .
  \end{array} \label{eq:impl}
\end{equation}
In other words, if the single vector $\nu^{\ast}$ dominates the set
$\mathcal{B}_{n_1 : n_2}$ w.r.t. realizations, then the set $\mathcal{A}_{n_1
: n_2}$ dominates the set $N_{n_1 : n_2} \cap \mathcal{C} (\delta)$ likewise.

The concentration statement {\tmem{GC}} in {\textsection}\ref{sec:intro}
says that given $\delta, \varepsilon, \eta > 0$, there is a number $\hat{c}
= \hat{c} (\delta, \varepsilon, \eta) \geqslant 1$ such that when the data
$b^E, b^I$ are scaled by any factor $c \geqslant \hat{c}$, then
\begin{equation}
  \nu^{\ast} \in \mathcal{A}_{n_1 : n_2} (\delta, \eta) \qquad \text{and}
  \qquad \frac{\# \nu^{\ast}}{\#\mathcal{B}_{n_1 : n_2} (\delta, \eta)} \;
  \geqslant \; \frac{1}{\varepsilon} . \label{eq:concGdiff}
\end{equation}
We establish the inequality in (\ref{eq:concGdiff}) by finding a lower bound
on $\# \nu^{\ast}$ in {\textsection}\ref{sec:lbNnu*} and an upper bound on $\#
\mathcal{B}_{n_1 : n_2} (\delta, \eta)$ in {\textsection}\ref{sec:ubNB}.
Theorem \ref{th:Gdiff} presents the ratio of these bounds. Then in
{\textsection}\ref{sec:scale} we find the concentration threshold $\hat{c}$
that ensures (\ref{eq:concGdiff}); that is given by Theorem \ref{th:cGdiff}.
Table \ref{tab:scaling} describes our notation for the process of scaling the
problem data.
\begin{table}[h]
  \centering
  \begin{tabular}{|c|c|}
    \hline
    Basic quantities & Derived quantities\\
    \hline
    $x^{\ast} \mapsto cx^{\ast}$ & $\nu^{\ast}$\\
    $s_1, s_2, s^{\ast} \mapsto cs_1, cs_2, cs^{\ast}$ & $n_1, n_2,
    n^{\ast}$\\
    $G^{\ast} \mapsto cG^{\ast}$ & \ \\
    $\vartheta_{\infty} \mapsto c \vartheta_{\infty}$ & \ \\
    \hline
  \end{tabular}
  \caption{\label{tab:scaling}\small The data scaling process $b \mapsto
    cb$. The symbols $x^{\ast}, s_1, s_2, \ldots$ on the left denote quantities
    before scaling. The symbols $\nu^{\ast}, \ldots$ on the right are quantities
    derived from the scaled basic quantities.}
\end{table}

\subsection{Realizations of the optimal count vector\label{sec:lbNnu*}}

In this section we find a lower bound on $\# \nu^{\ast} \; = \;
\binom{\nu^{\ast}_1 + \cdots + \nu^{\ast}_m}{\nu^{\ast}_1, \ldots,
\nu^{\ast}_m}$ where $\nu^{\ast}$ is the $m$-vector of Definition
\ref{def:nu*}, in terms of quantities related to the generalized entropy.

Like the number of realizations of a {\tmem{frequency}} vector and its
entropy, the number of realizations $\# \nu$ of a {\tmem{count}} vector $\nu$
is related to its generalized entropy. Given $\nu \in \mathbb{N}^m$, w.l.o.g.
let $\nu_1, \ldots, \nu_k, k \geqslant 1$ be its non-zero elements; then
\begin{equation}
  e^{- \frac{k}{12}} S (\nu) e^{G (\nu)} \leqslant \; \# \nu \; \leqslant S
  (\nu) e^{G (\nu)}, \qquad S (\nu) \; \triangleq \; \frac{\sqrt{n}}{(2
  \pi)^{(k - 1) / 2}}  \frac{1}{\sqrt{\nu_1 \cdots \nu_k}} . \label{eq:S}
\end{equation}
This follows immediately from eq. (III.6) in {\cite{entc2016}}, or Problem 2.2
in {\cite{CK2011}}; the bounds hold even when $k = 1$ and $\# \nu = 1$. Since
$\nu^{\ast}$ has no 0 elements (Remark \ref{rem:zeros}) we can take $k = m$ in
(\ref{eq:S}), so
\begin{equation}
  \# \nu^{\ast} \; \geqslant \; e^{- m / 12} S (\nu^{\ast}) e^{G (\nu^{\ast})}
  . \label{eq:numnu*1}
\end{equation}
Next we want to bound $G (\nu^{\ast})$ in terms of $G^{\ast} = G (x^{\ast})$.
By Proposition \ref{prop:nu*}, $\| \nu^{\ast} - x^{\ast} \|_{\infty} \leqslant
1$. If we assume that $x^{\ast} >\tmmathbf{1}$, Lemma \ref{le:close} applies
to $\nu^{\ast}$ and $x^{\ast}$ and we get
\begin{equation}
  x^{\ast} >\tmmathbf{1} \quad \Rightarrow \quad G (\nu^{\ast}) \; \geqslant
  \; G^{\ast} - \sum_i \ln \frac{1}{\chi^{\ast}_i} - \frac{1}{2}  \Bigl(
  \sum_i \frac{1}{x^{\ast}_i - 1} - \frac{m}{s^{\ast} / m - 1} \Bigr) .
  \label{eq:Gnu*lb}
\end{equation}
Returning to (\ref{eq:numnu*1}), it remains to find a convenient lower bound
for $S (\nu^{\ast})$. Since $\| \nu^{\ast} - x^{\ast} \|_{\infty} \leqslant
1$, we can use $\nu^{\ast}_i \leqslant x^{\ast}_i + 1$ in (\ref{eq:S}) to
obtain
\begin{equation}
  S (\nu^{\ast}) \; \geqslant \; \frac{\sqrt{s^{\ast}}}{(2 \pi)^{(m - 1) / 2}}
  \prod_{1 \leqslant i \leqslant m} \frac{1}{\sqrt{x^{\ast}_i + 1}} .
  \label{eq:Snulb}
\end{equation}
[Another, simpler bound, is obtained by noting that $\nu_1 \nu_2 \cdots \nu_m$
  is maximum when all $\nu_i$ are equal to $n / m$. The bound (\ref{eq:Snulb})
  is generally better, but can become slightly worse in some exceptional
  situations.] Putting (\ref{eq:Snulb}) and (\ref{eq:Gnu*lb}) in
(\ref{eq:numnu*1}),
\begin{equation}
  \begin{aligned}
    \# \nu^{\ast} & \geqslant \frac{e^{- m / 12} \sqrt{s^{\ast}}}{(2 \pi)^{(m -
        1) / 2}} e^{- \frac{1}{2} \Bigl( \sum_{i = 1}^m \frac{1}{x^{\ast}_i - 1}
      - \frac{m}{s^{\ast} / m - 1} \Bigr)} \prod_{i = 1}^m
    \frac{\chi^{\ast}_i}{\sqrt{x^{\ast}_i + 1}} \; \; e^{G^{\ast}}\\
    & \triangleq C_0 (x^{\ast}) e^{G^{\ast}}.
  \end{aligned}, \quad \text{if $x^{\ast} >\tmmathbf{1}$} \label{eq:numnu*2}
\end{equation}
The form of $C_0 (x^{\ast})$ is convenient for scaling according to Table
\ref{tab:scaling}.
\begin{remark}
  \label{rem:xs1}On the condition $x^{\ast} >\tmmathbf{1}$. It is certainly
  possible to formulate {\maxgent} problems whose solutions have some elements
  that are smaller than 1, in fact arbitrarily close to 0, and thus invalidate
  (\ref{eq:Gnu*lb}) and (\ref{eq:numnu*2}). Here however we are dealing with
  `large' problems, where $x^{\ast}$ is scaled by $c > 1$ for concentration to
  arise; see Theorem \ref{th:cGdiff} below. So one way to deal with such
  problem formulations is to take as ``the problem'' a certain pre-scaling of
  the original, one might say pathological, problem. Nevertheless, if one
  wanted to avoid the $x^{\ast} >\tmmathbf{1}$ issue entirely, one could use a
  weaker bound than (\ref{eq:Gnu*lb}) not subject to this restriction; see,
  for example, Remark \ref{rem:boundcomp}.
\end{remark}

\begin{remark}
  \label{rem:boundcomp}We compare the bound (\ref{eq:Gnu*lb}), derived for
  count vectors, to one adapted from a bound for density vectors. In
  {\cite{entc2016}}, proof of Proposition III.1, we derived the bound
  \begin{equation}
    H (f^{\ast}) \; \geqslant \; H (\chi^{\ast}) - \frac{3 m}{8 s^{\ast}} \ln
    (m - 1) - h \left( \frac{3 m}{8 s^{\ast}} \right) \label{eq:hdiffl1}
  \end{equation}
  where $h (\cdot)$ is the binary entropy function; there we had $n$ in place
  of $s^{\ast}$. [This is based on the bound $| H (\chi) - H (\psi) | \;
  \leqslant \; \frac{1}{2}  \| \chi - \psi \|_1 \ln (m - 1) + h \left(
  \frac{1}{2} \| \chi - \psi \|_1 \right)$; see {\cite{CK2011}} problem 3.10,
  or {\cite{Zha2007}}. An improved version, using both the $\ell_1$ and
  $\ell_{\infty}$ norms is in {\cite{Sason2013}}.] By multiplying both sides
  of (\ref{eq:hdiffl1}) by $n^{\ast}$ and then using the fact that $n^{\ast} H
  (f^{\ast}) = G (\nu^{\ast})$, $n^{\ast} \geqslant s^{\ast}$, and $s^{\ast} G
  (\chi^{\ast}) = G (x^{\ast}) \triangleq G^{\ast}$, we obtain
  \begin{equation}
    G (\nu^{\ast}) \; \geqslant \; G^{\ast} - \frac{3 m}{8} \ln (m - 1) -
    s^{\ast} h \left( \frac{3 m}{8 s^{\ast}} \right) . \label{eq:GbyHdiff}
  \end{equation}
  One way to compare the bounds (\ref{eq:Gnu*lb}) and (\ref{eq:GbyHdiff}) is
  to ask how the right-hand sides, apart from the $G^{\ast}$ term, behave
  under scaling of the problem by $c$ ({\textsection}\ref{sec:scaling}): we
  see that as $c$ increases, the r.h.s. of (\ref{eq:Gnu*lb}) tends to $-
  \sum_i \ln (1 / \chi^{\ast}_i)$ while the r.h.s. of (\ref{eq:GbyHdiff}) goes
  to $- \infty$.
\end{remark}

\subsection{Realizations of the sets with smaller entropy\label{sec:ubNB}}

Here we derive upper bounds on the number of realizations of the sets
$\mathcal{B}_n (\delta, \eta)$ and $\mathcal{B}_{n_1 : n_2} (\delta, \eta)$.
By combining them with the lower bound on $\# \nu^{\ast}$ of
{\textsection}\ref{sec:lbNnu*}, we establish our first main result, Theorem
\ref{th:Gdiff}.

From (\ref{eq:AnBneta}) and (\ref{eq:S}),
\[
   \# \mathcal{B}_n (\delta, \eta) \; \leqslant \; \sum_{\nu \in N_n \cap
   \mathcal{C} (\delta), \; G (\nu) < (1 - \eta) G^{\ast}} S (\nu) e^{G (\nu)}
   \quad \leqslant \quad e^{(1 - \eta) G^{\ast}}  \sum_{\nu \in N_n } S (\nu),
\]
where in going from the 1st to the 2nd inequality we ignored all the
constraints. Using (\ref{eq:S}) and proceeding as in {\cite{entc2016}}, proof
of Lemma III.1,
\begin{equation}
  \begin{aligned}
    \# \mathcal{B}_n (\delta, \eta) & \leqslant \; e^{(1 - \eta) G^{\ast}} 
    \sum_{k = 1}^m \binom{m}{k} \frac{\sqrt{n}}{(2 \pi)^{(k - 1) / 2}} 
    \sum_{\nu_1 + \cdots + \nu_k = n} \frac{1}{\sqrt{\nu_1 \cdots \nu_k}} \\
    & \leqslant \; e^{(1 - \eta) G^{\ast}}  \sum_{k = 1}^m \binom{m}{k} 
    \frac{\sqrt{n}}{(2 \pi)^{(k - 1) / 2}}  \int_{\tmscript{\begin{array}{c}
      x_1 + \cdots + x_k = n\\
      x_1, \ldots, x_k \geqslant 0
    \end{array}}} \frac{d x_1 \cdots d x_k}{\sqrt{x_1 \cdots x_k}} \\
    & =\; e^{(1 - \eta) G^{\ast}}  \sum_{k = 1}^m \binom{m}{k} 
    \frac{\sqrt{n}}{(2 \pi)^{(k - 1) / 2}}  \frac{\pi^{k / 2}}{\Gamma (k / 2)}
    n^{k / 2 - 1} \\
    & =\; e^{(1 - \eta) G^{\ast}}  \sqrt{2 \pi / n}  \sum_{k = 1}^m
    \binom{m}{k} \left( \frac{n}{2} \right)^{k / 2} \frac{1}{\Gamma (k / 2)}.
  \end{aligned} \label{eq:nBn1}
\end{equation}
We show in the Appendix {\textsection}\ref{app:pineq1} that the sum in the last
line above is bounded by $\frac{(n/2)^{m/2}}{\Gamma(m/2)} \bigl( 1 + \sqrt{m/n}
\bigr)^m$.  This is better than the $4 \bigl( 1 + \sqrt{n/4} \bigr)^m$ bound for
the same sum obtained in {\cite{entc2016}}, proof of Lemma III.1, as it is
asymptotically tight ($m$ fixed, $n \rightarrow \infty$). Using this improved
bound in (\ref{eq:nBn1}),
\begin{equation}
  \# \mathcal{B}_n (\delta, \eta) \; < \; e^{(1 - \eta) G^{\ast}}  \sqrt{2 \pi
  / n}  \frac{(n / 2)^{m / 2}}{\Gamma (m / 2)}  \left( 1 + \sqrt{m / n}
  \right)^m . \label{eq:nBn2}
\end{equation}
We now turn to the set $\mathcal{B}_{n_1 : n_2} (\delta, \eta)$ defined in
(\ref{eq:ABn1n2}). By (\ref{eq:nBn2}),
\[ \begin{aligned}
     \#\mathcal{B}_{n_1 : n_2} (\delta, \eta) & = \sum_{n_1 \leqslant n
     \leqslant n_2} \#\mathcal{B}_n (\delta, \eta) \\
     & < e^{(1 - \eta) G^{\ast}}  \frac{\sqrt{2 \pi}}{\Gamma (m / 2)} 
     \sum_{n_1 \leqslant n \leqslant n_2} \frac{1}{\sqrt{n}}  \left(
     \frac{n}{2} \right)^{\frac{m}{2}}  \left( 1 + \sqrt{m / n} \right)^m .
   \end{aligned} \]
Bounding the sum in the 2nd line by an integral,
\begin{align*}
     \frac{1}{2^{m / 2}}  \sum_{n_1 \leqslant n \leqslant n_2}
     \frac{1}{\sqrt{n}}  \left( \sqrt{n} + \sqrt{m} \right)^m & \leqslant
     \frac{1}{2^{m / 2}}  \int_{s_1}^{s_2 + 2} \frac{1}{\sqrt{y}}  \left(
     \sqrt{y} + \sqrt{m} \right)^m d y \\
     & = \frac{1}{2^{m / 2 - 1}}  \frac{1}{m + 1}  \Bigl( \bigl(\sqrt{s_2 + 2} +
     \sqrt{m} \bigr)^{m + 1} - \bigl(\sqrt{s_1} + \sqrt{m} \bigr)^{m+1} \Bigr)
\end{align*}
where in the first line we have widened the interval of integration from $[n_1,
  n_2 + 1]$ to $[s_1, s_2 + 2]$; recall the definition (\ref{eq:n1n2}) of $n_1,
n_2$.  Therefore
\begin{equation}
  \begin{gathered}
    \#\mathcal{B}_{n_1 : n_2} (\delta, \eta) \; < \; C_1 (s_1, s_2) e^{(1 -
    \eta) cG^{\ast}}, \\
    C_1 (s_1, s_2) \; \triangleq \; \frac{\sqrt{\pi}}{(m + 1) 2^{(m - 3) / 2}
    \Gamma (m / 2)} \left( \left( \sqrt{s_2 + 2} + \sqrt{m} \right)^{m + 1} -
    \left( \sqrt{s_1} + \sqrt{m} \right)^{m + 1} \right),
  \end{gathered} \label{eq:Bn1n2ref}
\end{equation}
where the sums $s_1, s_2$ have been defined in (\ref{eq:s1s2}).

By combining (\ref{eq:numnu*2}) and (\ref{eq:Bn1n2ref}) we arrive at our
first main result, a lower bound on the ratio of the number of realizations of
the optimal count vector $\nu^{\ast}$ to those of the set $\mathcal{B}_{n_1 :
n_2} (\delta, \eta)$, of count vectors with generalized entropy $\eta$-far
from $G^{\ast} = G (x^{\ast})$:

\begin{theorem}
  \label{th:Gdiff}Given structure matrices $A^E, A^I$ and data vectors $b^E,
  b^I$, let $(x^{\ast}, s_1, s_2)$ be the optimal solution to problem
  (\ref{eq:maxG}), (\ref{eq:s1s2}). Assume that $x^{\ast} >\tmmathbf{1}$;
  recall Remark \ref{rem:xs1}. Then for any $\delta, \eta > 0$,
  \[ \frac{\# \nu^{\ast}}{\#\mathcal{B}_{n_1 : n_2} (\delta, \eta)} \; > \;
     \frac{(m + 1) e^{- m / 12} \Gamma (m / 2)}{2 \pi^{\frac{m}{2}}} 
     \frac{C_2 (x^{\ast}) C_4 (x^{\ast})}{C_3 (s_1, s_2)} e^{\eta G^{\ast}} \]
  where the constants are
  \[ \begin{aligned}
       C_2 (x^{\ast}) & = \sqrt{s^{\ast}} \prod_{1 \leqslant i \leqslant m}
       \frac{\chi^{\ast}_i}{\sqrt{x^{\ast}_i + 1}}, \\
       C_4 (x^{\ast}) & = \mathrm{exp} \Bigl( - \frac{1}{2}  \Bigl( \sum_{1
       \leqslant i \leqslant m} \frac{1}{x^{\ast}_i - 1} - \frac{m}{s^{\ast} /
       m - 1} \Bigr) \Bigr), \\
       C_3 (s_1, s_2) & = \left( \sqrt{m} + \sqrt{s_2 + 2} \right)^{m + 1} -
       \left( \sqrt{m} + \sqrt{s_1} \right)^{m + 1} .
     \end{aligned} \]
\end{theorem}

One use of the theorem is when the problem is already `large' enough and doesn't
require further scaling. Then one may substitute appropriate values for $\delta$
and $\eta$ and see what kind of concentration is achieved. Note that the
concentration tolerance $\varepsilon$ does not appear in the theorem.

\subsection{The scaling factor needed for concentration\label{sec:scale}}

What happens to the lower bound of Theorem \ref{th:Gdiff} as the size of the
problem increases? In this section we establish Theorem \ref{th:cGdiff}, our
first concentration result, which shows that the bound can exceed $1 /
\varepsilon$ for any given $\varepsilon > 0$.

Introducing into the bound of Theorem \ref{th:Gdiff} a scaling factor $c
\geqslant 1$,
\begin{equation}
  \frac{\# \nu^{\ast}}{\#\mathcal{B}_{n_1 : n_2} (\delta, \eta)} \: > \;
  \frac{(m + 1) e^{- m / 12} \Gamma (m / 2)}{2 \pi^{\frac{m}{2}}}  \frac{C_2
  (cx^{\ast}) C_4 (cx^{\ast})}{C_3 (cs_1, cs_2)} e^{c \eta G^{\ast}} .
  \label{eq:cref0}
\end{equation}
To facilitate scaling, we develop bounds on the functions $C_2, C_3, C_4$ of
$c$ appearing above. First,
\begin{equation}
  C_2 (cx^{\ast}) \; = \; \frac{\sqrt{s^{\ast}}}{c^{(m - 1) / 2}}  \prod_{1
  \leqslant i \leqslant m} \frac{\chi^{\ast}_i}{\sqrt{x^{\ast}_i + 1 / c}} \;
  \geqslant \; \frac{\sqrt{s^{\ast}}}{c^{(m - 1) / 2}}  \prod_{1 \leqslant i
  \leqslant m} \frac{\chi^{\ast}_i}{\sqrt{x^{\ast}_i + 1}}, \quad c \geqslant
  1, \label{eq:C2b}
\end{equation}
since $\chi^{\ast}$ is invariant under scaling, and the first product above
increases as $c \nearrow$. Next, writing $C_3$ as
\[ C_3 (cs_1, cs_2) \; = \; c^{(m + 1) / 2}  \Bigl( \bigl( \sqrt{m/c} +
   \sqrt{s_2 + 2 / c} \bigr)^{m + 1} - \bigl( \sqrt{m / c} + \sqrt{s_1}
   \bigr)^{m + 1} \Bigr), \]
it can be shown that the function of $c$ multiplying $c^{(m + 1) / 2}$ above
decreases as $c \nearrow$\footnote{After some algebra, its derivative can be
shown to be negative if $s_2 \geqslant s_1$.}, so its maximum occurs at $c =
1$. Thus
\begin{equation}
  C_3 (cs_1, cs_2) \; \leqslant \; c^{(m + 1) / 2}  \left( \left( \sqrt{m} +
  \sqrt{s_2 + 2} \right)^{m + 1} - \left( \sqrt{m} + \sqrt{s_1} \right)^{m +
  1} \right), \quad c \geqslant 1. \label{eq:C3b}
\end{equation}
Finally, for $C_4 (cx^{\ast})$,
\[ - \frac{1}{2}  \Bigl( \sum_{1 \leqslant i \leqslant m} \frac{1}{cx^{\ast}_i
   - 1} - \frac{m}{cs^{\ast} / m - 1} \Bigr) \; > \; - \frac{1}{2}  \sum_{1
   \leqslant i \leqslant m} \frac{1}{cx^{\ast}_i - 1} \; \geqslant \; -
   \frac{1}{2}  \sum_{1 \leqslant i \leqslant m} \frac{1}{x^{\ast}_i - 1}, \]
since $x^{\ast} >\tmmathbf{1}$ and $c \geqslant 1$, and so
\begin{equation}
  C_4 (cx^{\ast}) \; > \; e^{- \frac{1}{2}  \sum_{i = 1}^m \frac{1}{x^{\ast}_i
  - 1}} . \label{eq:C4b}
\end{equation}
Putting (\ref{eq:C2b}), (\ref{eq:C3b}), and (\ref{eq:C4b}) into
(\ref{eq:cref0}), if $x^{\ast} >\tmmathbf{1}$,
\begin{equation}
  \frac{\# \nu^{\ast}}{\#\mathcal{B}_{n_1 : n_2} (\delta, \eta)} \; > \; Bc^{-
  m} e^{c \eta G^{\ast}}, \label{eq:cref1}
\end{equation}
where the constant
\[ B \; \triangleq \; \frac{(m + 1) \Gamma (m / 2) e^{- m / 12}}{2
   \pi^{\frac{m}{2}}}  \frac{\sqrt{s^{\ast}}  \prod_{1 \leqslant i \leqslant
   m} \frac{\chi^{\ast}_i}{\sqrt{x^{\ast}_i + 1}}}{\left( \sqrt{m} + \sqrt{s_2
   + 2} \right)^{m + 1} - \left( \sqrt{m} + \sqrt{s_1} \right)^{m + 1}} \; e^{-
  \frac{1}{2}  \sum_{i = 1}^m \frac{1}{x^{\ast}_i - 1}} \]
is $\ll \; 1$. By (\ref{eq:concGdiff}), the scaling factor $c$ to be applied
to the original problem must be such that the r.h.s. of (\ref{eq:cref1}) is
$\geqslant 1 / \varepsilon$, and also such that $\nu^{\ast}$ belongs to
$\mathcal{A}_{n_1 : n_2} (\delta, \eta)$. The first of these requirements
translates into
\begin{equation}
  c \eta G^{\ast} - m \ln c \; \geqslant \; - \ln (\varepsilon B),
  \label{eq:cref2}
\end{equation}
If $c_1$ is the largest of the two solutions of the equality version of
(\ref{eq:cref2})\footnote{An equation of this type generally has two roots,
one small and one large. For example $e^x / x = 10$ has roots 0.1118 and
3.577.}, the inequality (\ref{eq:cref2}) will hold for all $c \geqslant c_1$.

The second requirement on $c$, that $\nu^{\ast} \in \mathcal{A}_{n_1 : n_2}
(\delta, \eta)$, which is really $\nu^{\ast} \in \mathcal{A}_{n^{\ast}}
(\delta, \eta)$, has two parts. For the first part we need $\nu^{\ast} \in
\mathcal{C} (\delta)$; by Proposition \ref{prop:theta_inf} this is ensured by
$\| \nu^{\ast} - cx^{\ast} \|_{\infty} \leqslant \delta c \vartheta_{\infty}$,
and since the l.h.s. is $\leqslant 1$ by Proposition \ref{prop:nu*}, this will
hold if $c \geqslant c_2$ where
\begin{equation}
  c_2 \triangleq \frac{1}{\delta \vartheta_{\infty}} . \label{eq:cref3}
\end{equation}
For the second part we need $c$ to be s.t. $G (\nu^{\ast})
> (1 - \eta) cG^{\ast}$. By Proposition \ref{prop:x*scale} and
(\ref{eq:Gnu*lb}), this is ensured by
\begin{eqnarray}
  &  & cG^{\ast} - \sum_{1 \leqslant i \leqslant m} \ln
  \frac{1}{\chi^{\ast}_i} - \frac{1}{2}  \Bigl( \sum_{1 \leqslant i \leqslant
  m} \frac{1}{cx^{\ast}_i - 1} - \frac{m}{cs^{\ast} / m - 1} \Bigr) \; > \; (1
  - \eta) cG^{\ast} \nonumber\\
  &  & \Leftarrow \quad \frac{1}{2}  \sum_{1 \leqslant i \leqslant m}
  \frac{1}{cx^{\ast}_i - 1} + \sum_{1 \leqslant i \leqslant m} \ln
  \frac{1}{\chi^{\ast}_i} \; < \; c \eta G^{\ast}, \nonumber\\
  &  & \Leftarrow \quad \frac{1}{2 c}  \sum_{1 \leqslant i \leqslant m}
  \frac{1}{x^{\ast}_i - 1} + \sum_{1 \leqslant i \leqslant m} \ln
  \frac{1}{\chi^{\ast}_i} \; \leqslant \; c \eta G^{\ast},  \label{eq:cref4}
\end{eqnarray}
where the last implication follows from $c \geqslant 1$ and $x^{\ast}
>\tmmathbf{1}$. So we need $c \geqslant c_3$, the largest solution of the
(quadratic) equation version of (\ref{eq:cref4}).

\begin{figure}[h]
  \centering
  \resizebox{10.5cm}{!}{\includegraphics{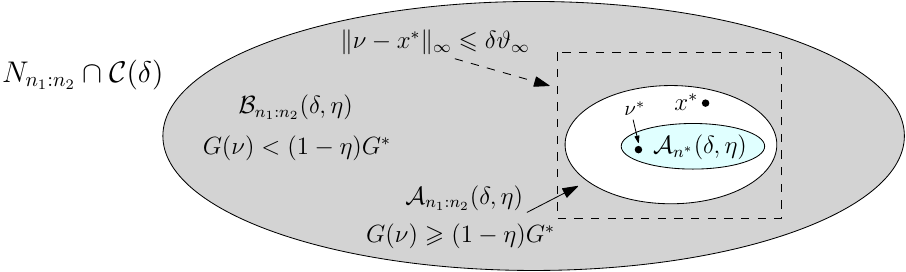}}
  \caption{\label{fig:cxstar-H}\small The outer ellipse, the set $N_{n_1 : n_2}
    \cap \mathcal{C} (\delta)$ of count vectors that satisfy the constraints to
    within tolerance $\delta$, is partitioned into $\mathcal{B}_{n_1 : n_2}$,
    shown in gray, and $\mathcal{A}_{n_1 : n_2}$, the inner white ellipse. The
    relationship shown between $\| \nu - x^{\ast} \|_{\infty} \leqslant \delta
    \vartheta_{\infty}$ and $\mathcal{A}_{n_1 : n_2} (\delta, \eta)$ is not the
    only one possible. Likewise for $x^{\ast}$ and $\mathcal{A}_{n_{}^{\ast}}
    (\delta, \eta)$.}
\end{figure}

Given tolerances $\delta, \varepsilon, \eta$, we have now established how to
compute a lower bound $\hat{c}$, the {\tmem{concentration threshold}}, on the
scaling factor required for concentration to occur around the point $\nu^{\ast}$
or in the set $\mathcal{A}_{n_1 : n_2}$, to the extent specified by $\delta,
\varepsilon, \eta$. This is our second main result, which establishes the
statement {\tmem{GC}} in {\textsection}\ref{sec:intro} concerning deviation from
the value $G^{\ast}$:
\begin{theorem}
  \label{th:cGdiff}With the conditions of Theorem \ref{th:Gdiff}, for any
  $\delta, \varepsilon, \eta > 0$, define the concentration threshold
  \[ \hat{c} \triangleq \max (c_1, c_2, c_3), \]
  where $c_1, c_2, c_3$ have been defined in
  (\ref{eq:cref2})-(\ref{eq:cref4}). Then when the data $b^E, b^I$ is scaled
  by a factor $c \geqslant \hat{c}$, the count vector $\nu^{\ast}$ of
  Definition \ref{def:nu*} belongs to the set $\mathcal{A}_{n_1 : n_2}
  (\delta, \eta)$ and we have
  \[ \frac{\# \nu^{\ast}}{\#\mathcal{B}_{n_1 : n_2} (\delta, \eta)} \geqslant
     \frac{1}{\varepsilon} \quad \text{and} \quad \frac{\#\mathcal{A}_{n_1 :
     n_2} (\delta, \eta)}{\# (N_{n_1 : n_2} \cap \mathcal{C} (\delta))}
     \geqslant 1 - \varepsilon, \]
  where $n_1 = \lceil cs_1 \rceil, n_2 = \lceil cs_2 \rceil$, and the sets
  $\mathcal{A}_{n_1 : n_2}, \mathcal{B}_{n_1 : n_2}$ have been defined in
  (\ref{eq:ABn1n2}).
\end{theorem}
Note that the constraint information $A^E, b^E, A^I, b^I$ appears implicitly,
via $s_1, s_2$, and $\vartheta_{\infty}$. The various sets figuring in the
theorem are depicted in Figure \ref{fig:cxstar-H}.

\subsubsection{Bounds on the concentration threshold\label{sec:bcth}}

It is useful to know something about how the threshold $\hat{c}$ depends on
the solution $x^{\ast}, G^{\ast}$ to the {\maxgent} problem and on the
parameters $\delta, \varepsilon, \eta$, without having to solve equations. We
derive some bounds on $\hat{c}$ with regard to convenience, not
tightness\footnote{The bounds {\tmem{still}} require knowing the solution
$x^{\ast}$ to the {\maxgent} problem.}.

If $c_i \geqslant L_i$, then $\hat{c} = \max_i c_i \geqslant \max_i L_i$.
Hence we have the lower bound
\begin{equation}
  \hat{c} \; \geqslant \; \max \Bigl( \frac{- \ln (\varepsilon B)}{\eta
  G^{\ast}}, \frac{1}{\delta \vartheta_{\infty}} \Bigr), \label{eq:chatlb}
\end{equation}
since $c_1$ must be bigger than the first term on the r.h.s., and $c_2$ equals
the second. As intuitively expected, the bound says that the smaller $\delta,
\varepsilon$, or $\eta$ are, the more scaling we need. By looking at the
expression for $B$ after (\ref{eq:cref1}), we see that the same holds the
farther apart the bounds $s_1, s_2$ on the possible sums are from each other;
this accords with intuition, and we discuss it further in Example
\ref{ex:scale}.

Next, if $c_i \leqslant U_i$, then $\max_i c_i \leqslant \max_i U_i$. So
\begin{equation}
  \hat{c} \; \leqslant \; \max \biggl( \frac{2 m}{\eta G^{\ast}} \ln \frac{m -
  \ln (\varepsilon B)}{\eta G^{\ast}} - \frac{\ln (\varepsilon B)}{\eta
  G^{\ast}}, \frac{1}{\delta \vartheta_{\infty}}, \sqrt{\frac{\sum_{i = 1}^m 1
  / (x^{\ast}_i - 1)}{2 \eta G^{\ast}}} \biggr), \label{eq:chatub}
\end{equation}
where the expressions on the r.h.s. are upper bounds on $c_1, c_2, c_3$,
respectively, as shown in the Appendix\footnote{Concerning the last
expression, recall our assumption $x^{\ast} >\tmmathbf{1}$ and Remark
\ref{rem:xs1}.}. The upper bound says that the larger $G^{\ast}$ is, the less
scaling we need; likewise for the elements of $x^{\ast}$. Both of these
implications agree with intuition. Further illustrations of the bounds
(\ref{eq:chatlb}) and (\ref{eq:chatub}) are in Example \ref{ex:imp2}.

\subsection{Examples}

We give two examples. The first continues Example \ref{ex:imp}, illustrates
the bounds on the concentration threshold, and points out a, at first sight,
surprising behavior of the threshold. The second example illustrates an
intuitively-expected relationship between concentration and the bounds $s_1,
s_2$.

\begin{example}
  \label{ex:imp2}Returning to Example \ref{ex:imp}, we find
  \[ s_1 = 21.5, \; s_2 = 37.5, \; x^{\ast} = (6.591, 5.326, 13.26, 1.120,
     2.253, 2.789), \; s^{\ast} = 31.34, \; G^{\ast} = 47.53. \]
  Thus $\nu^{\ast} = (7, 5, 14, 1, 2, 3)$ and $n^{\ast} = 32$. Also,
  $\vartheta_{\infty} = 2.9$. Table \ref{tab:exfirst} shows what happens when
  the problem data $b$ is scaled by the factor $\hat{c}$ dictated by the given
  $\delta, \varepsilon, \eta$. [We don't use a special notation for the
  quantities appearing in the unscaled vs. the scaled problem, so whenever we
  write $x^{\ast}, \nu^{\ast}$, $b^E$, etc. a scaling factor, which could be
  1, is implied.]
  
  \begin{table}[h]
    \centering
    \footnotesize
    \begin{tabular}{|c|c|c|ccc|c|} \hline
      \multicolumn{7}{|c|}{$\delta \in [0.01, 1], \varepsilon = 10^{- 9}$} \\
      \hline
      $\eta$ & $\hat{c}$ & $b$ & $n_1$ & $n^{\ast}$ & $n_2$ & $\nu^{\ast}$\\
      \hline
      0.05 & 34.48 & (362.1,631.0,300.0,137.9) & 880 & 1081 & 1294 &
      (227,184,457,39,78,96)\\
      0.02 & 91.27 & (958.3,1670,794.0,365.1) & 2328 & 2861 & 3423 &
      (602,486,1210,102,206,255)\\
      0.01 & 191.9 & (2015,3512,1670,767.7) & 4894 & 6015 & 7197 &
      (1265,1022,2545,215,433,535)\\
      \hline
      \multicolumn{7}{|c|}{$\delta \in [0.01, 1], \varepsilon = 10^{- 15}$} \\
      \hline
      0.05 & 40.25 & (422.7,736.6,350.2,161.0) & 1027 & 1262 & 1510 &
      (266,214,534,45,91,112)\\
      0.02 & 106.8 & (1121.3,1954.3,929.1,427.2) & 2724 & 3347 & 4005 &
      (703,569,1416,120,241,298)\\
      0.01 & 222.9 & (2340.2,4078.6,1939.0,891.5) & 5684 & 6985 & 8358 &
      (1469,1187,2955,250,502,622) \\
      \hline
    \end{tabular}
    \caption{\label{tab:exfirst}Scaling of the problem of Example \ref{ex:imp}
    for the given $\delta, \varepsilon, \eta$.}
  \end{table}
  
  With respect to the discrete solution, in the first row of Table
  \ref{tab:exfirst} for example, we have $\| x^{\ast} - \nu^{\ast} \|_{\infty}
  = 0.370$. Further, $\nu^{\ast}$ satisfies the equality constraints with
  tolerance $\| A^E \nu^{\ast} - b^E \|_{\infty} / \min | b^E | = 0.0033$ and
  the inequality constraints with tolerance 0. We see that the scaling factor
  $\hat{c}$ is quite sensitive to $\eta$ and rather insensitive to
  $\varepsilon$; this can be surmised from (\ref{eq:chatlb}). One way to
  interpret the scaling is as a change in the {\tmem{scale of measurement}} of
  the data $b$, e.g. a change in the units. Then scaling by a larger factor
  means choosing more refined units, and the above results show that the
  concentration increases, as intuitively expected.
  
  With respect to the bounds (\ref{eq:chatlb}) and (\ref{eq:chatub}) on the
  threshold $\hat{c}$, for the first row of the table with $\delta = 0.01$,
  they yield $\hat{c} \in [34.48, 41.87]$. For $\delta \in [0.02, 0.05]$ they
  yield $\hat{c} \in [25.1, 41.87]$. For the second row, the bounds give
  $\hat{c} \in [62.8, 116.2]$ for any $\delta \in [0.01, 0.05]$.
  
  Now suppose that the problem data is {\tmem{pre-scaled}} by 34.5. Then for
  the first row the bounds say that $\hat{c} \in [1.0, 1.0]$, i.e. no further
  scaling is needed. For the second row, Theorem \ref{th:cGdiff} gives
  $\hat{c} = 2.39$ and the bounds give $\hat{c} \in [2.23, 2.55]$. So the
  original problem had a threshold $\hat{c} = 91.27$, but when scaled by 34.5,
  the threshold becomes only $\hat{c} = 2.39 < \frac{91.27}{34.5} = 2.64$.
  Apparently, unlike the rest of the problem (Proposition \ref{prop:x*scale}),
  the concentration threshold does not behave linearly with scaling: $\hat{c}
  \left( 34.5 \times \text{problem} \right) < 34.5 \hat{c} \left(
  \text{problem} \right)$. The explanation for this at first sight
  disconcerting behavior is two-fold: first, Theorem \ref{th:cGdiff} does not
  say that $\hat{c}$ is the \tmtextit{minimum} required scaling factor for a
  given problem; second, there are many approximations involved in the
  derivation of $\hat{c}$, and many get better as the size of the problem
  increases.
\end{example}

\begin{example}
  \label{ex:scale}Intuition says that the bounds $s_1, s_2$ on the possible
  sums of the admissible count vectors have something to do with
  concentration: if they are wide, concentration should be more difficult to
  achieve. Suppose that, somehow, the {\maxgent} vector $x^{\ast}$ from which
  $\nu^{\ast}$ is derived remains fixed; then the wider the range $s_1, s_2$
  allowed by the constraints, the larger should be the scaling factor required
  for $\nu^{\ast}$ to dominate. The bound (\ref{eq:chatlb}) agrees with this,
  due to the expression for $B$ after (\ref{eq:cref1}). We now give a simple
  situation in which the difference between $s_1$ and $s_2$ can increase while
  $x^{\ast}$ remains fixed.
  
  Consider a 2-dimensional problem with box constraints $b_1 \leqslant x_1
  \leqslant b_2$, $b_3 \leqslant x_2 \leqslant b_4$, depicted in Fig.
  \ref{fig:scale}. Then $s_1 = b_1 + b_3$, $s_2 = b_2 + b_4$ and $G$ is
  maximum at the upper right corner of the box (Proposition \ref{prop:max}).
  If we reduce $b_1, b_3$ to $b'_1, b'_3$, the lower left corner of the box
  moves down and to the left while the upper right corner remains fixed, as
  shown in the figure.
  \begin{figure}[h]
    \centering
    \resizebox{6cm}{!}{\includegraphics{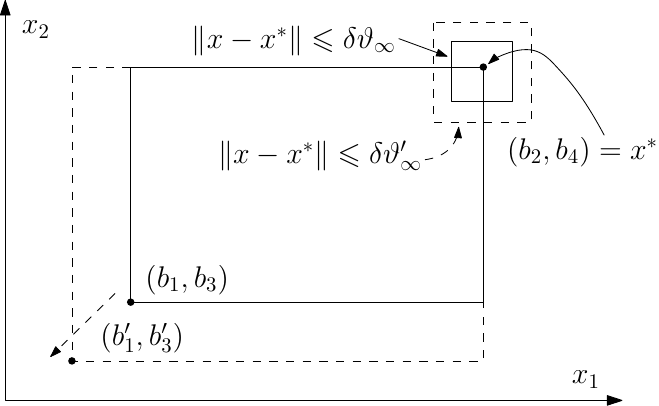}}
    \caption{\label{fig:scale}\small Reducing $s_1$ while leaving $s_2$ and
    $x^{\ast}$ unchanged.}
  \end{figure}
  Thus we widen the bounds $s_1, s_2$ while leaving $s^{\ast}, G^{\ast}$
  unchanged, and the problem with the new box constraints requires more
  scaling than the original problem. The construction generalizes immediately
  to $m$ dimensions, see {\textsection}\ref{sec:lb}.
\end{example}

\section{Concentration with respect to distance from the {\maxgent}
vector\label{sec:dconc}}

In this section we provide results analogous to those of
{\textsection}\ref{sec:gconc}, but with the sets $\mathcal{A}, \mathcal{B}$
formulated in terms of the {\tmem{distance}} of their elements from the optimal
{\tmem{vector}} $x^{\ast}$, as measured by the $\ell_1$ norm. This is a more
intuitive measure than difference in entropy. There are three main results:
Theorems \ref{th:cdist1} and \ref{th:cdist2}, analogues of Theorems
\ref{th:Gdiff} and \ref{th:cGdiff}, and Theorem \ref{th:cdist3}, an optimized
version of Theorem \ref{th:cdist2} that does not require specifying a
$\delta$. In various places we reuse results and methods from
{\textsection}\ref{sec:gconc}, so the presentation here is more succinct.

For given $n$ and $\delta > 0$, we want to consider the count vectors in $N_n$
that lie in $\mathcal{C} (\delta)$ and whose distance from $x^{\ast}$ is no
more than $\vartheta > 0$ in $\ell_1$ norm, and those that lie in $\mathcal{C}
(\delta)$ but are farther from $x^{\ast}$ than $\vartheta$ in $\ell_1$ norm.
The situation is less straightforward than with frequency/density vectors.
First, given two real $m$-vectors, the norm of their difference can never be
smaller than the difference of their norms, so it does not make sense to
require that this norm be too small\footnote{In the case of frequency vectors,
this lower bound is 0. See Proposition \ref{prop:far} for more details.}.
Second, we will be considering norms that can be large numbers, especially
after scaling of the problem, so it will not do to consider a fixed-size
region around $x^{\ast}$. For these reasons, we define for $\vartheta > 0$
\begin{equation}
  \begin{gathered}
    \mathcal{A}_n (\delta, \vartheta) \; \triangleq \; \{\nu \in N_n \cap
    \mathcal{C}(\delta), \| \nu - x^{\ast} \|_1 \leqslant | n - s^{\ast} | +
    \min (n, s^{\ast}) \vartheta\},\\
    \mathcal{B}_n (\delta, \vartheta) \; \triangleq \; \{\nu \in N_n \cap
    \mathcal{C}(\delta), \| \nu - x^{\ast} \|_1 > | n - s^{\ast} | + \min (n,
    s^{\ast}) \vartheta\} .
  \end{gathered} \label{eq:AnBntheta}
\end{equation}
This is more complicated that the definition for frequency vectors in
{\cite{entc2016}}, but here $\vartheta$ is again a small number $< \; 1$. If
$n$ were equal to $s^{\ast}$, (\ref{eq:AnBntheta}) would say that the density
vectors $f$ and $\chi^{\ast}$ are such that $\| f - \chi^{\ast} \|_1$ is
$\leqslant \; \vartheta$ in $\mathcal{A}_n$ and $> \; \vartheta$ in
$\mathcal{B}_n$. In general, (\ref{eq:AnBntheta}) says that the norm of $\nu -
x^{\ast}$ is close to $| n - s^{\ast} |$: if $n \leqslant s^{\ast}$, the bound
is $s^{\ast} - (1 - \vartheta) n$, and if $n > s^{\ast}$ it is $n - (1 -
\vartheta) s^{\ast}$.

We will consider the (disjoint) unions of the sets (\ref{eq:AnBntheta}) over
$n \in \{ n_1, \ldots, n_2 \}$, with $n_1, n_2$ given by (\ref{eq:n1n2}):
\begin{equation}
  \begin{gathered}
    \mathcal{A}_{n_1 : n_2} (\delta, \vartheta) \; \triangleq \; \bigl\{ \nu
    \mid \sum_i \nu_i = n, n_1 \leqslant n \leqslant n_2, \nu \in
    \mathcal{C} (\delta), \| \nu - x^{\ast} \|_1 \leqslant | n - s^{\ast} | +
    \min (n, s^{\ast}) \vartheta \bigr\}, \\
    \mathcal{B}_{n_1 : n_2} (\delta, \vartheta) \; \triangleq \; \bigl\{ \nu
    \mid \sum_i \nu_i = n, n_1 \leqslant n \leqslant n_2, \nu \in
    \mathcal{C} (\delta), \| \nu - x^{\ast} \|_1 > | n - s^{\ast} | + \min (n,
    s^{\ast}) \vartheta \bigr\} .
  \end{gathered} \label{eq:ABn1n2n}
\end{equation}
For any $\delta, \vartheta$, these two sets partition $N_{n_1 : n_2} \cap
\mathcal{C} (\delta)$, the set of count vectors that sum to a number between
$n_1$ and $n_2$ and lie in $\mathcal{C} (\delta)$.

With these definitions, we will establish an analogue of (\ref{eq:concGdiff})
in {\textsection}\ref{sec:gconc}: given $\delta, \varepsilon, \vartheta > 0$,
there is a concentration threshold $\hat{c} = \hat{c} (\delta, \varepsilon,
\vartheta)$ s.t. if the problem data $b^E, b^I$ is scaled by any factor $c
\geqslant \hat{c}$, then the {\maxgent} count vector $\nu^{\ast}$ is in the
set $\mathcal{A}_{n_1 : n_2} (\delta, \vartheta)$ and has at least $1 /
\varepsilon$ times the realizations of all vectors in the set
$\mathcal{B}_{n_1 : n_2} (\delta, \vartheta)$:
\begin{equation}
  \nu^{\ast} \in \mathcal{A}_{n_1 : n_2} (\delta, \vartheta) \qquad \text{and}
  \qquad \frac{\# \nu^{\ast}}{\#\mathcal{B}_{n_1 : n_2} (\delta, \vartheta)}
  \; \geqslant \; \frac{1}{\varepsilon} . \label{eq:NnuBntheta}
\end{equation}
There is one important difference with {\textsection}\ref{sec:gconc}, that
here the tolerances $\delta$ and $\vartheta$ cannot be chosen
{\tmem{independently}} of one another, they must obey a certain restriction.

\begin{remark}
  \label{rem:delta}\tmcolor{red}{} $G^{\ast}$ is the maximum of $G$ over the
  domain $\mathcal{C} (0)$, with no tolerances on the constraints. As we said
  in {\textsection}\ref{sec:constr}, a tolerance $\delta > 0$ widens this
  domain to $\mathcal{C} (\delta)$, may move the vector that maximizes $G$
  from $x^{\ast} (0)$ to $x^{\ast} (\delta)$, and may change the maximum value
  from $G^{\ast} (0)$ to $G^{\ast} (\delta)$. Here we are looking for
  concentration in a region of size $\vartheta$ around the point $x^{\ast}$.
  If $\delta$ is too large, we cannot expect such a region to dominate the
  count vectors in $\mathcal{C} (\delta)$ w.r.t. the number of realizations,
  since $x^{\ast} (\delta)$ may even lie inside the set $\mathcal{B} (\delta,
  \vartheta)$; by Proposition \ref{prop:max}, it already lies on the
  `boundary' of $\mathcal{C} (0)$. If $\vartheta$ is given, concentration in
  $\mathcal{A} (\delta, \vartheta)$ requires an upper bound on the allowable
  $\delta$; see (\ref{eq:dt}) below.
  
  In the setting of {\textsection}\ref{sec:gconc} there is no limitation on
  the magnitude of $\delta$ with respect to that of $\eta$. It is perfectly
  fine if the set $\mathcal{A}_n (\delta, \eta)$ contains $\nu$ with $G (\nu)
  > G^{\ast} (0)$, but not if $\mathcal{B}_n (\delta, \eta)$ does. But
  $\mathcal{B}_n (\delta, \eta)$ can't contain any such $\nu$ \tmtextit{by its
  definition} (\ref{eq:AnBneta}): if there are any such $\nu$, all of them
  have to be in $\mathcal{A}_n (\delta, \eta)$.
\end{remark}

\subsection{Realizations of the sets far from the {\maxgent}
vector\label{sec:far}}

To bound the number of realizations of $\mathcal{B}_{n_1 : n_2} (\delta,
\vartheta)$ we need to show that if $\nu$ is far from $x^{\ast}$, in the $\|
\nu - x^{\ast} \|_1$ sense, then $G (\nu)$ is far from $G (x^{\ast})$. To
simplify the notation, in this section we denote $x^{\ast} (0), \chi^{\ast}
(0), G^{\ast} (0)$ simply by $x^{\ast}, \chi^{\ast}, G^{\ast}$.

We first need an auxiliary relationship between the norm of the difference of
two real vectors and the norm of the difference of their normalized versions:
\begin{proposition}
  \label{prop:far}Let $\| \cdot \|$ be any vector norm, such as $\| \cdot
  \|_1, \| \cdot \|_2, \| \cdot \|_{\infty}$ etc. Then for any $x, y \in
  \mathbb{R}^m$ and $\vartheta > 0$,
  \[ \| x - y \| > | \| x \| - \| y \| | + \vartheta \quad \Rightarrow \quad
     \left\| \frac{x}{\| x \|} - \frac{y}{\| y \|} \right\| >
     \frac{\vartheta}{\min (\| x \|, \| y \|)} . \]
\end{proposition}
What we want to show about $G (\nu)$ and $G^{\ast}$ follows by taking Lemma
\ref{le:far}, bounding the divergence term $D (\cdot \| \cdot)$ in terms of
the $\ell_1$ norm, and then using Proposition \ref{prop:far} with the $\ell_1$
norm and $\min (\| \nu \|, \| x^{\ast} \|) \vartheta$ in place of $\vartheta$:
\begin{lemma}
  \label{le:far2}Given $\delta \geqslant 0$ and $\vartheta > 0$, with the
  notation of Lemma \ref{le:far}, for any count vector $\nu \in \mathcal{C}
  (\delta)$ with sum $n$,
  \[ \| \nu - x^{\ast} \|_1 > | n - s^{\ast} | + \min (n, s^{\ast}) \vartheta
     \quad \Rightarrow \quad G (\nu) \; \leqslant \; G^{\ast} + \Lambda^{\ast}
     \delta - \gamma^{\ast} \vartheta^2 n \]
  where
  \[ \gamma^{\ast} \triangleq \frac{1}{4 (1 - 2 \beta^{\ast})} \ln \frac{1 -
     \beta^{\ast}}{\beta^{\ast}}, \quad \beta^{\ast} \triangleq \max_{I
     \subset \{1, \ldots, m\}} \min \Bigl( \sum_{i \in I} \chi^{\ast}_i, 1 -
     \sum_{i \in I} \chi^{\ast}_i \Bigr) . \]
  In general, $\gamma^{\ast} \geqslant \frac{1}{2}$ and $\frac{1 -
  \chi^{\ast}_{\max}}{2} \leqslant \beta^{\ast} \leqslant \frac{1}{2}$. If
  $\beta^{\ast} = 1 / 2$, $\gamma^{\ast} \triangleq 1 / 2$.
\end{lemma}

The bound on the divergence that we used above, $D (p\|q) \geqslant \gamma (q)
\| p - q \|_1^2$, is due to {\cite{pinsker2}}. The closeness of the number
$\beta (q)$ to $1 / 2$ can be thought of as measuring how far away the density
vector $q$ is from having a partition\footnote{In the sense of the NP-complete
problem {\tmname{Partition}}.}. {\cite{BHK2014}} is also relevant here, as the
authors study $\inf_p D (p\|q)$ subject to $\| p - q \|_1 \geqslant \ell$.
They refer to $1 - \beta \geqslant 1 / 2$, where $\beta$ is as in Lemma
\ref{le:far2}, as the ``balance coefficient''. Their Theorem 1b provides an
exact value for $\inf_p D (p\|q)$ as a function of $1 - \beta$, $q$, and
$\ell$, valid for $\ell \leqslant 4 (1 / 2 - \beta)$; this could be used in
Lemma \ref{le:far2}, at the expense of an additional condition between, in our
notation, $\vartheta$ and $\beta^{\ast}$. They also show that $\beta \geqslant
1 / 2 - q_{\max} / 2$, where $q_{\max}$ is the largest element of $q$, a
result which we have incorporated into Lemma \ref{le:far2}.

We can now proceed to find an upper bound on $\# \mathcal{B}_{n_1 : n_2}
(\delta, \vartheta)$. Beginning with $\#\mathcal{B}_n (\delta, \vartheta)$, by
(\ref{eq:AnBntheta}) and (\ref{eq:S})
\[ \# \mathcal{B}_n (\delta, \vartheta) \; \leqslant
   \sum_{\tmscript{\begin{array}{c}
     \nu \in N_n \cap \mathcal{C} (\delta)\\
     \| \nu - x^{\ast} \|_1 > | n - s^{\ast} | + \min (n, s^{\ast}) \vartheta
   \end{array}}} \hspace{-2em} S (\nu) e^{G (\nu)} . \]
Applying Lemma \ref{le:far2} to $G (\nu)$ and, similarly to what we did in
{\textsection}\ref{sec:ubNB}, ignoring the condition involving the norm in the
sum as well as the intersection with $\mathcal{C} (\delta)$,
\[ \# \mathcal{B}_n (\delta, \vartheta) \; \leqslant \; e^{G^{\ast} +
   \Lambda^{\ast} \delta - \gamma^{\ast} \vartheta^2 n}  \; \sum_{\nu \in N_n}
   S (\nu) . \]
The sum above is identical to that in the expression for $\#\mathcal{B}_n
(\delta, \eta)$ given at the beginning of {\textsection}\ref{sec:ubNB}, so
following the development that led to (\ref{eq:nBn2}),
\[ \# \mathcal{B}_n (\delta, \vartheta) \; \leqslant \; \sqrt{2 \pi / n} \;
   \frac{(n / 2)^{m / 2}}{\Gamma (m / 2)}  \bigl( 1 + \sqrt{m / n} \bigr)^m
   e^{G^{\ast} + \Lambda^{\ast} \delta - \gamma^{\ast} \vartheta^2 n} . \]
Compare with (\ref{eq:nBn2}). Consequently,
\begin{equation}
  \begin{aligned}
    & \# \mathcal{B}_{n_1 : n_2} (\delta, \vartheta) \; = \; \sum_{n_1 \leqslant
    n \leqslant n_2} \# \mathcal{B}_n (\delta, \vartheta) \\
    & \qquad \leqslant \; \frac{\sqrt{2 \pi}}{2^{m/2} \Gamma (m/2)} \;
    e^{G^{\ast} + \Lambda^{\ast} \delta}  \sum_{n_1 \leqslant n \leqslant n_2}
    \frac{1}{\sqrt{n}} \bigl(\sqrt{m} + \sqrt{n}\bigr)^m  \, e^{-\gamma^{\ast} \vartheta^2 n} \\
    & \qquad \leqslant \; \frac{\sqrt{2 \pi}}{2^{m/2} \Gamma (m / 2)} \,
    e^{G^{\ast} + \Lambda^{\ast} \delta} \\
    & \hspace{4em} \frac{2}{(m+1)}  \Bigl( \bigl(\sqrt{s_2 + 2} + \sqrt{m} \bigr)^{m+1} \,
    e^{- \gamma^{\ast} \vartheta^2  (s^{\ast} + 1)} + \bigl(
    \sqrt{s^{\ast} + 2} + \sqrt{m} \bigr)^{m + 1} \, e^{-\gamma^{\ast} \vartheta^2 s_1} \Bigr)
  \end{aligned} \label{eq:numBn1n2bound}
\end{equation}
where the inequality implied in the last line is derived in the Appendix. This
bound on $\# \mathcal{B}_{n_1 : n_2} (\delta, \vartheta)$ is to be compared
with the bound (\ref{eq:Bn1n2ref}) on $\# \mathcal{B}_{n_1 : n_2} (\delta,
\eta)$.

Combining (\ref{eq:numnu*2}) with (\ref{eq:numBn1n2bound}) we obtain a lower
bound on the ratio of numbers of realizations analogous to that of Theorem
\ref{th:Gdiff}:
\begin{theorem}
  \label{th:cdist1}Given structure matrices $A^E, A^I$ and data vectors $b^E,
  b^I$, let $(x^{\ast}, s_1, s_2)$ be the optimal solution of problem
  (\ref{eq:maxG}). Assume that $x^{\ast} >\tmmathbf{1}$; recall Remark
  \ref{rem:xs1}. Then for any $\delta, \varepsilon, \vartheta > 0$,
  \[ \frac{\# \nu^{\ast}}{\# \mathcal{B}_{n_1 : n_2} (\delta, \vartheta)} \; >
     \; \frac{(m + 1) \Gamma (m / 2) e^{- m / 12}}{2 \pi^{m / 2}}  \frac{C_2
     (x^{\ast}) C_4 (x^{\ast})}{C'_3 (s^{\ast}, s_1, s_2)} \;  \;
     e^{\gamma^{\ast} \vartheta^2 s_1 - \Lambda^{\ast} \delta}, \]
  where the constants $C_2 (x^{\ast}), C_4 (x^{\ast})$ are the same as in
  (\ref{eq:cref0}),
  \[ C'_3 (s^{\ast}, s_1, s_2) \; = \; \left( \sqrt{s_2 + 2} + \sqrt{m}
     \right)^{m + 1} e^{- \gamma^{\ast} \vartheta^2  (s^{\ast} - s_1 + 1)} + \left(
     \sqrt{s^{\ast} + 2} + \sqrt{m} \right)^{m + 1}, \]
  and $\Lambda^{\ast}, \gamma^{\ast}$ have been defined in Lemmas \ref{le:far}
  and \ref{le:far2} respectively.
\end{theorem}

The lower bound will not be useful if the exponent $\gamma^{\ast} \vartheta^2
s_1 - \Lambda^{\ast} \delta$ is not positive. We elaborate on this in
{\textsection}\ref{sec:cdist}. Also, like Theorem \ref{th:Gdiff}, the theorem
says nothing about {\tmem{how large}} the bound is for a given problem. This
is the job of Theorem \ref{th:cdist2}.

\subsection{Scaling and concentration around the {\maxgent} count
vector\label{sec:cdist}}

As we did in {\textsection}\ref{sec:scale}, we now investigate what happens to
the lower bound of Theorem \ref{th:cdist1} when the problem data $b$ is scaled
by a factor $c \geqslant 1$. The end results are the concentration Theorems
\ref{th:cdist2} and \ref{th:cdist3} below.

Table \ref{tab:scaling} described how scaling the data affects the quantities
appearing in the bound, except for $\Lambda^{\ast}$, which is new to
{\textsection}\ref{sec:dconc}. Scaling $b$ has the effect $x^{\ast} \mapsto
cx^{\ast}$, and from (\ref{eq:x*j}) in {\textsection}\ref{sec:max} we see that
the Lagrange multipliers remain unchanged\footnote{This also follows from
expression (\ref{eq:G*lambda}) in the proof of Lemma \ref{le:far}, for
$G^{\ast}$ in terms of the multipliers and $G^{\ast} \mapsto cG^{\ast}$.}.
Then the definition of $\Lambda^{\ast}$ in Lemma \ref{le:far} shows that the
end result of scaling is $\Lambda^{\ast} \mapsto c \Lambda^{\ast}$. This and
$s_1 \mapsto cs_1$ imply that scaling by $c$ multiplies the exponent
$\gamma^{\ast} \vartheta^2 s_1 - \Lambda^{\ast} \delta$ in Theorem
\ref{th:cdist1} by $c$. The effect of scaling on $C_2$ and $C_4$ is given by
(\ref{eq:C2b}) and (\ref{eq:C4b}), and finally
\begin{equation}
  \begin{array}{l}
    C'_3 (cs^{\ast}, cs_1, cs_2) \; < \; \\
    \qquad c^{\frac{m + 1}{2}}  \bigl((\sqrt{s_2 + 2} + \sqrt{m})^{m+1} \,
    e^{-\gamma^{\ast} \vartheta^2} e^{- c \gamma^{\ast}
      \vartheta^2 (s^{\ast} - s_1)} + (\sqrt{s^{\ast} + 2} + \sqrt{m})^{m+1}
     \bigr) \\
     \qquad \leqslant \; c^{\frac{m + 1}{2}} \bigl( (\sqrt{s_2 + 2} +
     \sqrt{m})^{m + 1} e^{- \gamma^{\ast} \vartheta^2 (s^{\ast} - s_1 + 1)} +
     (\sqrt{s^{\ast} + 2} + \sqrt{m})^{m + 1} \bigr)\\
    \qquad \triangleq \; c^{\frac{m + 1}{2}} C''_3,
  \end{array} \label{eq:C3pp}
\end{equation}
where the 2nd inequality follows from $c \geqslant 1$.
In conclusion, when the data $b$ is scaled by the factor $c \geqslant 1$,
Theorem \ref{th:cdist1} says that if $x^{\ast} >\tmmathbf{1}$, then
\begin{equation}
  \frac{\# \nu^{\ast}}{\# \mathcal{B}_{n_1 : n_2} (\delta, \vartheta)} \; > \;
  \frac{B'}{C''_3} c^{- m} e^{(2 \gamma^{\ast} \vartheta^2 s_1 -
  \Lambda^{\ast} \delta) c}, \label{eq:nlb1}
\end{equation}
where $C''_3$ is defined in (\ref{eq:C3pp}) and
\begin{equation}
  B' \; \triangleq \; \frac{(m + 1) \Gamma (m / 2) e^{- m / 12}}{2 \pi^{m /
  2}}  \sqrt{s^{\ast}} e^{- \frac{1}{2}  \sum_{i = 1}^m \frac{1}{x^{\ast}_i -
  1}}  \prod_{1 \leqslant i \leqslant m} \frac{\chi^{\ast}_i}{\sqrt{x^{\ast}_i
  + 1}} . \label{eq:nlb2}
\end{equation}
(\ref{eq:nlb1}) and (\ref{eq:nlb2}) are to be compared with (\ref{eq:cref1}).

Recalling Remark \ref{rem:delta}, an important consequence of (\ref{eq:nlb1})
is that if concentration is to occur the tolerances $\delta$ and $\vartheta$
must satisfy
\begin{equation}
  \vartheta^2 \; > \; \frac{\Lambda^{\ast}}{2 \gamma^{\ast} s_1} \delta .
  \label{eq:dt}
\end{equation}
This can be ensured by choosing small enough $\delta$ for the given
$\vartheta$, or large enough $\vartheta$ for the given $\delta$. (The results
of this paper do not immediately translate to the frequency vector case, but
(\ref{eq:dt}) can be compared with the similar condition in Theorem IV.2 of
{\cite{entc2016}}.)

By (\ref{eq:nlb1}), the concentration statement (\ref{eq:NnuBntheta}) will
hold if the scaling factor $c$ is such that
\begin{equation}
  (2 \gamma^{\ast} \vartheta^2 s_1 - \Lambda^{\ast} \delta) c - m \ln c \;
  \geqslant \; \ln \frac{C''_3}{\varepsilon B'} . \label{eq:nlb3}
\end{equation}
This inequality is of the same form as (\ref{eq:cref2}), and will hold for all
$c$ greater than the larger of the two roots of the equality version of it.

As in {\textsection}\ref{sec:scale}, we also need $\nu^{\ast}$ to be in the
set $\mathcal{A}_{n_1 : n_2} (\delta, \vartheta)$ of (\ref{eq:ABn1n2n}), more
specifically in $\mathcal{A}_{n^{\ast}} (\delta, \vartheta)$. For this, we
must first have $\nu^{\ast} \in \mathcal{C} (\delta)$; this is ensured by $c
\geqslant c_2$, with $c_2$ as in (\ref{eq:cref3}). Second, by the definition
(\ref{eq:AnBntheta}) of $\mathcal{A}_n (\delta, \vartheta)$, we need $\|
\nu^{\ast} - x^{\ast} \|_1 \leqslant | n^{\ast} - s^{\ast} | + \min (n^{\ast},
s^{\ast}) \vartheta$; by Proposition \ref{prop:nu*} this will hold if
$\vartheta \geqslant (3 m / 4 + 1) / (cs^{\ast})$.

We have now established the desired analogue of Theorem \ref{th:cGdiff}, and
proved the statement {\tmem{GC}} of {\textsection}\ref{sec:intro}, in terms of
distance from the {\maxgent} vector $x^{\ast}$:
\begin{theorem}
  \label{th:cdist2}With the same conditions as in Theorem \ref{th:cdist1},
  suppose that the tolerances $\delta, \vartheta$ satisfy (\ref{eq:dt}), where
  $\Lambda^{\ast}, \gamma^{\ast}$ have been defined in Lemmas \ref{le:far} and
  \ref{le:far2}. Let
  \[ c_1 \; \triangleq \; \frac{3 m / 4 + 1}{\vartheta s^{\ast}}, \quad c_2 \;
     \triangleq \; \frac{1}{\delta \vartheta_{\infty}}, \]
  and given $\varepsilon > 0$, let $c_3$ be the largest root $c$ of the
  equality version of (\ref{eq:nlb3}). Finally, define the concentration
  threshold
  \[ \hat{c} \; \triangleq \; \max (c_1, c_2, c_3) . \]
  Then when the data $b^E, b^I$ is scaled by any $c \geqslant \hat{c}$, the
  {\maxgent} count vector $\nu^{\ast}$ of Definition \ref{def:nu*} belongs to
  the set $\mathcal{A}_{n_1 : n_2} (\delta, \vartheta)$ of (\ref{eq:ABn1n2n}),
  specifically to $\mathcal{A}_{n^{\ast}} (\delta, \vartheta)$, and is such
  that
  \[ \frac{\# \nu^{\ast}}{\#\mathcal{B}_{n_1 : n_2} (\delta, \vartheta)}
     \geqslant \frac{1}{\varepsilon} \quad \text{and} \quad
     \frac{\#\mathcal{A}_{n_1 : n_2} (\delta, \vartheta)}{\# (N_{n_1 : n_2}
     \cap \mathcal{C} (\delta))} > 1 - \varepsilon . \]
\end{theorem}

The second inequality in the claim of the theorem follows from the first by
(\ref{eq:impl}) in {\textsection}\ref{sec:gconc}, which holds whether the sets
$\mathcal{A}_{n_1 : n_2}$ and $\mathcal{B}_{n_1 : n_2}$ are defined as they
were in {\textsection}\ref{sec:gconc} or as they were defined here. As in
Theorem \ref{th:cGdiff}, the constraint information $A^E, b^E, A^I, b^I$
appears implicitly in Theorem \ref{th:cdist2}, via $s_1, s_2$, and
$\vartheta_{\infty}$. Bounds on the concentration threshold can be derived
similarly to {\textsection}\ref{sec:bcth}.

\begin{figure}[h]
  \centering
  \resizebox{10cm}{!}{\includegraphics{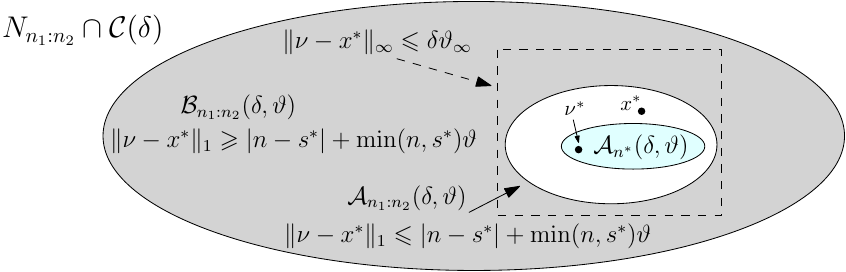}}
  \caption{\label{fig:cxstar-norm}\small Concentration around $\nu^{\ast}$
    w.r.t.  $\ell_1$ norm. The {\maxgent} vector $\nu^{\ast}$ has $1 /
    \varepsilon$ times more realizations than the entire set $\mathcal{B}_{n_1 :
      n_2} (\delta, \vartheta)$, shown in gray. The relationship we show between
    $\| \nu - x^{\ast} \|_{\infty} \leqslant \delta \vartheta_{\infty}$ and
    $\mathcal{A}_{n_1 : n_2} (\delta, \vartheta)$ is not the only one possible;
    likewise for $x^{\ast}$ and $\mathcal{A}_{n^{\ast}} (\delta, \vartheta)$.}
\end{figure}

Finally, Fig. \ref{fig:cxstar-norm} depicts the various sets involved in the
definition of the threshold $\hat{c}$ appearing in the theorem.

From the definition of $\hat{c}$ in Theorem \ref{th:cdist2} we see that as
$\delta$ increases, the constants $c_2$ and $c_3$ behave in opposite ways:
$c_2$ decreases but $c_3$ increases. If one cares only about the tolerances
$\varepsilon$ and $\vartheta$, and does not care to specify a particular
$\delta$, this opens the possibility of reducing $\hat{c}$ by choosing
$\delta$ so as to minimize the largest of $c_2, c_3$:

\begin{theorem}
  \label{th:cdist3}Given $\varepsilon, \vartheta$, suppose that
  \[ \Lambda^{\ast} \geqslant m \vartheta_{\infty} \quad \text{and} \quad
     \vartheta^2 \; < \; \frac{\Lambda^{\ast}  \sqrt{s^{\ast} / m +
     1}}{\gamma^{\ast} \vartheta_{\infty} s_1}  \frac{1}{\varepsilon^{1 / m}},
  \]
  where the various quantities are as in Theorem \ref{th:cdist2}. If so, the
  equation for $\delta$
  \[ \frac{2 \gamma^{\ast} \vartheta^2 s_1}{\vartheta_{\infty}} 
     \frac{1}{\delta} + m \ln \delta \; = \; \ln \frac{C''_3}{\varepsilon B'}
     + \frac{\Lambda^{\ast}}{\vartheta_{\infty}} - m \ln \vartheta_{\infty} \]
  has a root $\delta_0 \in \bigl(0, 2 \gamma^{\ast} \vartheta^2 s_1 /
  \Lambda^{\ast}\bigr]$, and we define
  \[ \hat{c} \; \triangleq \; \max \Bigl( \frac{1}{\delta_0
     \vartheta_{\infty}}, \frac{3 m / 4 + 1}{\vartheta s^{\ast}} \Bigr) . \]
  Then when the data $b^E, b^I$ is scaled by any $c \geqslant \hat{c}$, the
  {\maxgent} count vector $\nu^{\ast}$ of Definition \ref{def:nu*} belongs to
  the set $\mathcal{A}_{n^{\ast}} (\delta_0, \vartheta)$ of
  (\ref{eq:AnBntheta}), and is such that
  \[ \frac{\# \nu^{\ast}}{\#\mathcal{B}_{n_1 : n_2} (\delta_0, \vartheta)}
     \geqslant \frac{1}{\varepsilon} \quad \text{and} \quad
     \frac{\#\mathcal{A}_{n_1 : n_2} (\delta_0, \vartheta)}{\# (N_{n_1 : n_2}
     \cap \mathcal{C} (\delta_0))} > 1 - \varepsilon . \]
\end{theorem}

In this situation a simple lower bound on the concentration threshold
$\hat{c}$ is
\begin{equation}
  \hat{c} \; \geqslant \; \max \left( \frac{H (\chi^{\ast})}{2 \gamma^{\ast}} 
  \frac{1}{\vartheta_{\infty} \vartheta^2}, \frac{3 m}{4 s^{\ast} \vartheta}
  \right) \label{eq:chatlb3}
\end{equation}
where for the first expression we used the upper bound on $\delta_0$ and
$\Lambda^{\ast} \geqslant s^{\ast} H (\chi^{\ast})$. The ratio $H
(\chi^{\ast}) / \gamma^{\ast}$ is small for imbalanced distributions
$\chi^{\ast}$, e.g. with a single dominant element, in which case
$\gamma^{\ast}$ is large, and approaches $2 \ln m$ for perfectly balanced
ones. The bound (\ref{eq:chatlb3}) says that $\hat{c}$ increases with
$\vartheta^{- 2}$, and this can be seen in Example \ref{ex:3d} below.

\subsection{Examples\label{sec:ex2}}

The first two examples illustrate Theorems \ref{th:cdist2} and
\ref{th:cdist3}, while the third illustrates the removal of 0s from the
solution mentioned in {\textsection}\ref{sec:max} and the `boundary' case in
which the {\maxgent} vector $x^{\ast}$ sums to the maximum allowable $s^{\ast}
= s_2$.

\begin{example}
  \label{ex:2nd}We return to Example \ref{ex:imp2}. Recall that
  \[ s_1 = 21.5, \; s_2 = 37.5, \; x^{\ast} = (6.591, 5.326, 13.26, 1.120,
     2.253, 2.789), \; s^{\ast} = 31.34, \; G^{\ast} = 47.53. \]
  We have $\vartheta_{\infty} = 2.9$, $\gamma^{\ast} = 0.5$, and
  $\Lambda^{\ast} = G^{\ast} = 47.53$. The constraint (\ref{eq:dt}) on
  $\delta$ and $\vartheta$ is $\vartheta^2 > 1.864 \delta$. This means that if
  we want small $\vartheta$, we must have a correspondingly small $\delta$, as
  we commented after (\ref{eq:dt}). Table \ref{tab:ex2nd} lists various values
  of $\hat{c} (\delta, \varepsilon, \vartheta)$ obtained from Theorem
  \ref{th:cdist2}.
  
  \begin{table}[h]
    \centering
    \begin{tabular}{l}
      {\small{\begin{tabular}{|cc|cl|} \hline
        &  & \multicolumn{2}{c|}{$\hat{c}$} \\
        $\delta$ & $\vartheta$ & $\varepsilon = 10^{- 9}$ & $\varepsilon =
        10^{- 15}$\\  \hline
        0.001 & 0.08 & 862.7 & 989.2\\
        $10^{- 4}$ &  & 3448 & 3448\\
        $10^{- 5}$ &  & 34483 & 34483\\
        0.001 & 0.07 & 1322 & 1511\\
        $10^{- 4}$ &  & 3448 & 3448\\
        $10^{- 5}$ &  & 34483 & 34483\\
        0.001 & 0.06 & 2392 & 2722\\
        $10^{- 4}$ &  & 3448 & 3448\\
        $10^{- 5}$ &  & 34483 & 34483\\
        $10^{- 4}$ & 0.05 & 3448 & 3448\\
        $10^{- 5}$ &  & 34483 & 34483\\
        $10^{- 5}$ & 0.03 & 34483 & 34483\\
        & 0.01 & $60376$ & $67351$\\
        & 0.008 & $111472$ & $123967$ \\ \hline
      \end{tabular}}}
    \end{tabular}
    \caption{\label{tab:ex2nd}\small Scaling of the problem of Example
      \ref{ex:imp2} for the given $\delta, \varepsilon, \vartheta$. The
      threshold $\hat{c}$ does not behave smoothly because of the $\max ()$ in
      Theorem \ref{th:cdist2}.}
  \end{table}
\end{example}

\begin{example}
  \label{ex:3d}Consider the same data as in Table \ref{tab:ex2nd}, but with
  only $\varepsilon, \vartheta$ specified; we don't care about a particular
  $\delta$, as long as it ensures that $\nu^{\ast} \in \mathcal{A}_{n^{\ast}}
  (\delta, \vartheta)$. With $\delta$ chosen automatically by Theorem
  \ref{th:cdist3}, Table \ref{tab:ex3d} below shows that the concentration
  threshold $\hat{c}$ is significantly reduced.
  
  \begin{table}[h]
    \centering
    {\small{\begin{tabular}{|l|cc|} \hline
      & \multicolumn{2}{c|}{$\hat{c}$} \\
      $\vartheta$ & $\varepsilon = 10^{- 9}$ & $\varepsilon = 10^{- 15}$\\
      \hline
      0.08 & 704.4 & 793.4\\
      0.07 & 933.5 & 1050\\
      0.06 & 1292 & 1450\\
      0.05 & 1896 & 2124\\
      0.04 & 3032 & 3387\\
      0.03 & 5548 & 6178\\
      0.01 & $55345$ & 60991\\
      0.008 & 88189 & 97004 \\ \hline
    \end{tabular}}}
    \caption{\label{tab:ex3d}\small The threshold $\hat{c}$ for given
      $\varepsilon, \vartheta$ with optimal selection of $\delta =
      \delta_0$. Compare with Table \ref{tab:ex2nd}.}
  \end{table}
  
  The variation of $\hat{c}$ with $\vartheta^{- 2}$ implied by the lower bound
  (\ref{eq:chatlb3}) is evident.
\end{example}

\begin{example}
  \label{ex:vt}Fig. \ref{fig:vt} shows four cities connected by road segments.
  We assume that vehicles travelling from one city to another follow the most
  direct route, and that there is no traffic from a city to itself.
  
  \begin{figure}[h]
    \centering
    \resizebox{7cm}{!}{\includegraphics{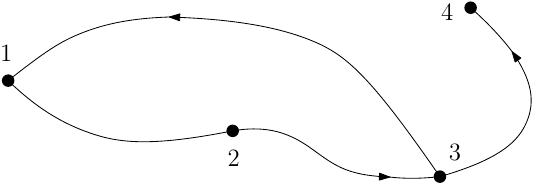}}
    \caption{\label{fig:vt}\small Four cities connected by (bidirectional) road
    segments. Arrows indicate the constrained directions.}
  \end{figure}
  
  The number of vehicles in city $i$ is known, which puts upper bounds on the
  number that leaves each city; also, from observations we have lower bounds on
  the number of vehicles on the road segments $2 \rightarrow 3$, $3 \rightarrow
  1$, and $3 \rightarrow 4$. From this information we want to infer how many
  vehicles travel from city $i$ to city $j$, i.e. infer the $4 \times 4$ matrix
  of counts
  \[ v = \left(\begin{array}{cccc}
       0 & v_{12} & v_{13} & v_{14}\\
       v_{21} & 0 & v_{23} & v_{24}\\
       v_{31} & v_{32} & 0 & v_{34}\\
       v_{41} & v_{42} & v_{43} & 0
     \end{array}\right) . \]
  So suppose the constraints on $v$ are
  \[ v_{i i} = 0, \quad \sum_j v_{i j} \leqslant 100, 120, 80, 90, \quad
     v_{23} + v_{24} \geqslant 80, \quad v_{31} + v_{41} \geqslant 59, \quad
     v_{14} + v_{24} + v_{34} \geqslant 70, \]
  where the last three reflect the ``direct route'' assumption. Then we have
  $s_1 = 139, s_2 = 390$. We define the 12-element vector $x$ for the
  {\maxgent} method as $(v_{12}, v_{13}, v_{14}, v_{21}, v_{23}, v_{24}, \allowbreak
  \ldots, v_{43})$. [Note that if we knew that \tmtextit{all} vehicles in a
  city leave the city, then we could define a {\tmem{frequency}} matrix by
  dividing the matrix $v$ by $100 + \cdots + 90$ and thus formulate a
  {\maxent} problem.]  The {\maxgent} solution is
  \[ v^{\ast} \; = \; \left(\begin{array}{cccc}
       0 & 33.333 & 33.333 & 33.333\\
       40.0 & 0 & 40.0 & 40.0\\
       27.765 & 26.118 & 0 & 26.118\\
       31.235 & 29.382 & 29.382 & 0
     \end{array}\right) \]
  with sum $s^{\ast} = s_2 = 390$ and maximum generalized entropy $G^{\ast} =
  964.62$, $\Lambda^{\ast}=971.84$, $\gamma^{\ast} = 0.5$. So here we have the
  boundary case in which the sum of $x^{\ast}$ is the maximum possible.
  (Problems involving matrices subject to constraints of the above type, for
  which {\tmem{analytical}} solutions are possible, were studied in
  {\cite{Oik2010}}.)
  
  Applying Theorem \ref{th:cdist3} with $\vartheta = 0.04, \varepsilon = 10^{-
    15}$ the `optimal' $\delta$ is $\delta_0 \approx 2.02 \cdot 10^{- 5}$ and
  yields the threshold $\hat{c} = 837.9$. Using a scaling factor $c = 838$ on
  $v^{\ast}$ results in the integral matrix
  \[ \nu^{\ast} \; = \; \left(\begin{array}{cccc}
     0 & 27394 & 27393 & 27393\\
     33520 & 0 & 33520 & 33520\\
     23267 & 21887 & 0 & 21887\\
     \mathrm{26175} & 24622 & 24622 & 0
   \end{array}\right) \]
  with sum $n^{\ast} = 326820 \in [116842, 326820]$.  This matrix has at least
  $10^{15}$ times the number of realizations of the entire set
  $\mathcal{B}_{116842 : 326820} (2.02 \cdot 10^{- 5}, 0.04)$ defined in
  (\ref{eq:ABn1n2n}). To gain some appreciation of what this means, it is not
  easy to determine the size of this set, but just the particular subset of it
  \[ \mathcal{B}_{326819} (0, 0.04) = \{ \nu \in \mathcal{C} (0), \| \nu -
  x^{\ast} \|_1 > 13073 \} \] contains at least $2.012 \cdot 10^{37}$
  elements\footnote{We have $| 326819 - 838 \cdot 390 | + \min (326819, 838
    \cdot 390) \cdot 0.04 \; = \; 13073.8.$}.  For comparison, the whole of
  $\mathcal{C} (0)$ has $2.394 \cdot 10^{54}$ elements.  (We compute
  these numbers with the \tmverbatim{barvinok} software, {\cite{Ver2005}}. For
  $\mathcal{B}_{326819}$ we get a lower bound by using the stronger constraint
  $| \nu_1 - x^{\ast}_1 | > 13072$ in place of $\| \nu - x^{\ast} \|_1 > 13073$,
  which is harder to express.)
\end{example}

\section{Conclusion\label{sec:concl}}

We demonstrated an extension of the phenomenon of entropy concentration,
hitherto known to apply to probability or frequency vectors, to the realm of
count vectors, whose elements are natural numbers. This required introducing a
new entropy function in which the sum of the count vector plays a role. Still,
like the Shannon entropy, this generalized entropy can be viewed combinatorially
as an approximation to the log of a multinomial coefficient.  Our derivations
are carried out in a fully discrete, finite, non-asymptotic framework, do not
involve any probabilities, and all of the objects about which we make any claims
are fully constructible. This discrete, combinatorial setting is an attempt to
reduce the phenomenon of entropy concentration to its essence. We believe that
this concentration phenomenon supports viewing the maximization of our
generalized entropy as a compatible extension of the well-known {\maxent} method
of inference.

\subsubsection*{Acknowledgments}

Thanks to Peter Gr{\"u}nwald for his comments on a previous version of the
manuscript, and for many useful discussions on the subject.

\appendix\section{Proofs}

\subsubsection*{Proof of Proposition \ref{prop:inc}}

Given a $y \geqslant x$, $y$ can be reached from $x$ by a sequence of steps
each of which increases a single coordinate, and the value of $G$ increases at
each step because all its partial derivatives are positive. (The derivatives
are 0 only at points $x$ that consist of a single non-zero element; a direct
proof can be given for that case.)

For a more formal proof, we note that the directional derivative $G' (\xi ;
u)$ of $G$ at any point $\xi$ is $\geqslant \; 0$ in any direction $u
\geqslant 0$: $G' (\xi ; u) = \nabla G (\xi) \cdot u = \sum_i u_i \ln
\frac{\xi_1 + \cdots + \xi_m}{\xi_i}$. So any move away from $\xi$ in a
direction $u \geqslant 0$ will increase $G$. More precisely, by the mean value
theorem, for any $y$ that can be written as $x + u$ for some $u \geqslant 0$,
there is a $\xi$ on the line segment from $x$ to $x + u$ s.t. $G (x + u) - G
(x) = \nabla G (\xi) \cdot u \geqslant 0$. Finally, if some element of $u$ is
strictly positive, then $\nabla G (\xi) \cdot u > 0$.

\subsubsection*{Proof of Proposition \ref{prop:Gconc}}

\begin{enumerate}
  \item To establish concavity it suffices to show that $\nabla^2 G (x)$, the
  Hessian of $G$, is negative semi-definite. We find
  \begin{equation}
    \nabla^2 G (x) \; = \; \frac{1}{x_1 + \cdots + x_m} U_m - \tmop{diag}
    \Bigl( \frac{1}{x_1}, \ldots, \frac{1}{x_m} \Bigr), \label{eq:hessian}
  \end{equation}
  where $U_m$ is a $m \times m$ matrix all of whose entries are 1. Given $x$,
  for an arbitrary $y = (y_1, \ldots, y_m)$ we must have $y^T \nabla^2 G (x) y
  \leqslant 0$. To show this, first write $\nabla^2 G (x)$ as
  \[ \nabla^2 G (x) \; = \; \frac{1}{x_1 + \cdots + x_m}  \Bigl( U_m -
     \tmop{diag} \Bigl( \frac{x_1 + \cdots + x_m}{x_1}, \ldots, \frac{x_1 +
     \cdots + x_m}{x_m} \Bigr) \Bigr) . \]
  Now define $\xi_i = x_i / (x_1 + \cdots + x_m)$. Then $y^T \nabla^2 G (x) y
  \leqslant 0$ is equivalent to
  \begin{equation}
    (y_1 + \cdots + y_m)^2 \; \leqslant \; y_1^2 / \xi_1 + \cdots + y_m^2 /
    \xi_m, \label{eq:yconv}
  \end{equation}
  where the $\xi_i$ are $\geqslant 0$ and sum to 1. But for fixed $y$, $y^2_1
  / \xi_1 + \cdots + y^2_m / \xi_m$ is a convex function of $\xi = (\xi_1,
  \ldots, \xi_m)$ over the domain $\xi \succ 0$, and its minimum under the
  constraint $\xi_1 + \cdots + \xi_m = 1$ occurs at $\xi_i = y_i / (y_1 +
  \cdots + y_m)$. So the least value of the r.h.s. of (\ref{eq:yconv}) as a
  function of $\xi_1, \ldots, \xi_m$ is $(y_1 + \cdots + y_m)^2$, and this
  establishes (\ref{eq:yconv}).
  
  For a given $x$ we see that $y^T \nabla^2 G (x) y$ is 0 exactly at points
  $y$ such that for all $i$, $x_i / \sum_j x_j = y_i / \sum_j y_j$, i.e. iff
  $y = cx$ for some $c \in \mathbb{R}$.
  
  \item The fact that the Hessian fails to be negative definite does not imply
  that $G$ is not strictly concave; negative definiteness is a sufficient, but
  not a necessary condition for strict concavity.
  
  It can be seen that $G$ is not strictly concave because of the scaling or
  homogeneity property \ref{Gpropsc} in {\textsection}\ref{sec:Gbasic}:
  consider the distinct points $x$ and $y = 2 x$; strict concavity would
  require $G ((x + y) / 2) > G (x) / 2 + G (y) / 2$, which is not true.
  
  \item Proposition 1.1.2 in Chapter IV of {\cite{HUL1996}} says that a
  function $F (x)$ is strongly convex on a convex set $C$ with modulus $\gamma
  > 0$ iff the modified function $F (x) - \frac{1}{2} \gamma \|x\|_2^2$ is
  convex on $C$. Applying this to our function $G$, by the proof carried out
  in part 1, we would have to show that given any $x \in \mathbb{R}_+^m$, for
  all $y \in \mathbb{R}_+^m$
  \[ (y_1 + \cdots + y_m)^2 - (y^2_1 / \xi_1 + \cdots + y^2_m / \xi_m) +
     \gamma (y^2_1 + \cdots + y^2_m) \; \leqslant \; 0 \]
  for the chosen modulus $\gamma > 0$. But for any $x$ and any $\gamma > 0$,
  this condition is false at the point $y = \gamma x$.
  
  \item By Definition 1.1.1 in {\cite{HUL1996}} Ch. V, {\textsection}1.1, a
  convex and positively homogeneous function $F$ defined over the extended
  real numbers $\mathbb{R} \cup \{\pm \infty\}$ is sublinear. If we define $G
  (\cdummy)$ over all of $\mathbb{R}^m$ by setting $G (x_1, \ldots, x_m) = -
  \infty$ if any $x_i$ is negative, the above statement applies to $F = - G$.
  Finally, a sublinear function has the property $F (\alpha x + \beta y)
  \leqslant \alpha F (x) + \beta F (y)$.
\end{enumerate}

\subsubsection*{Proof of Lemma \ref{le:close}}

By (\ref{eq:hcube}), if $\|x - y\|_{\infty} \leqslant \zeta$ we have $G (y)
\geqslant G (x -\tmmathbf{\zeta})$. Now we expand $G (x -\tmmathbf{\zeta})$ in
a Taylor series around $x$. Since $G (\cdummy)$ is a twice-differentiable
function on the open set $x > \tmmathbf{\zeta}$, if $x, x'$ are two points
in this set, then there is a $\tilde{x} = (1 - \alpha) x + \alpha x'$ with
$\alpha \in [0, 1]$, such that
\[ G (x') \; = \; G (x) + \nabla G (x) \cdot (x' - x) + \frac{1}{2}  (x' -
   x)^T \cdot \nabla^2 G (\tilde{x}) \cdot (x' - x) \]
(Theorem 12.14 of {\cite{Apo74}}). Set $x' = x -\tmmathbf{\zeta}$, so
$\tilde{x} = x - \alpha \zeta$. Noting that
\begin{align*}
     \nabla G (x) & = \Bigl( \ln \frac{x_1 + \cdots + x_m}{x_1}, \ldots, \ln
     \frac{x_1 + \cdots + x_m}{x_m} \Bigr),\\
     \nabla^2 G (\tilde{x}) & = \frac{1}{\tilde{x}_1 + \cdots + \tilde{x}_m}
     U_m - \mathrm{diag} \Bigl( \frac{1}{\tilde{x}_1}, \ldots,
     \frac{1}{\tilde{x}_m} \Bigr),\\
     (x' - x)^T \cdot \nabla^2 G (\tilde{x}) \cdot (x' - x) & = \zeta^2 
     \sum_i \Bigl( \frac{m^2}{\tilde{x}_1 + \cdots + \tilde{x}_m} -
     \frac{1}{\tilde{x}_i} \Bigr),
\end{align*}
where the second equality is (\ref{eq:hessian}) in the proof of Proposition
\ref{prop:Gconc}, we find that for any $x > \tmmathbf{\zeta}$
\begin{equation}
  G (x -\tmmathbf{\zeta}) \; = \; G (x) - \zeta \sum_i \ln \frac{x_1 + \cdots
  + x_m}{x_i} - \frac{1}{2} \zeta^2  \Bigl( \sum_i \frac{1}{x_i - \alpha
  \zeta} - \frac{m}{\|x\|_1 / m - \alpha \zeta} \Bigr), \label{eq:Gzeta1}
\end{equation}
where we know that the sum of the $\zeta$ and the $\zeta^2$ terms on the right
is negative. [We chose to expand around the point $x -\tmmathbf{\zeta}$
because then the sign of the terms $\nabla G (x) \cdot \tmmathbf{\zeta}$ and
$\tmmathbf{\zeta}^T \cdot \nabla^2 G (\tilde{x}) \cdot \tmmathbf{\zeta}$ is
known.] Now for fixed $x$ define the function
\begin{equation}
  g (\alpha, \zeta) \; \triangleq \sum_{1 \leqslant i \leqslant m}
  \frac{1}{x_i - \alpha \zeta} - \frac{m}{\|x\|_1 / m - \alpha \zeta}, \quad
  \alpha \in [0, 1], \zeta \geqslant 0, \; x > \tmmathbf{\zeta}.
  \label{eq:galpha}
\end{equation}
This function is $\geqslant \; 0$ and increasing with $\alpha$. To see that $g
(\alpha, \zeta) \geqslant 0$, set $u_i = x_i - \alpha \zeta$ so that $\|x\|_1
/ m - \alpha \zeta$ becomes the arithmetic mean $\bar{u}$ of the $u_i$; then
use a fundamental property of the power means: for any $u \geqslant 0$, and
any weights $w_i$ summing to 1,
\begin{equation}
  \biggl( \sum_i w_i u_i^{- k} \biggr)^{- 1 / k} \; \leqslant \; \sum_i w_i
  u_i, \qquad k \geqslant 1 \label{eq:pmeans}
\end{equation}
(see {\cite{HLP}}, Theorem 16). The desired result follows by choosing all
$w_i = 1 / m$. To show that $g (\alpha, \zeta)$ increases with $\alpha$,
\[ \frac{\partial g}{\partial \alpha} = \sum_i \frac{\zeta}{(x_i - \alpha
   \zeta)^2} - \frac{m \zeta}{(\|x\|_1 / m - \alpha \zeta)^2} \]
and this is always $\geqslant \; 0$ by the same power means technique
(\ref{eq:pmeans}). [Similarly, $\partial^2 g / \partial \alpha^2 \geqslant 0$,
so $g (\alpha, \zeta)$ is a convex function of $\alpha$.] We therefore see
that for any $\zeta \geqslant 0$
\begin{equation}
  \min_{\alpha \in [0, 1]} g (\alpha, \zeta) = g (0, \zeta) \qquad \text{and}
  \qquad \max_{\alpha \in [0, 1]} g (\alpha, \zeta) = g (1, \zeta) .
  \label{eq:gminmax}
\end{equation}
It now follows from $G (y) \geqslant G (x -\tmmathbf{\zeta})$,
(\ref{eq:Gzeta1}), and (\ref{eq:gminmax}) that for any $x >\tmmathbf{\zeta}$
and any $y$ s.t. $\|y - x\|_{\infty} \leqslant \zeta$
\[ G (y) \; \geqslant \; G (x) - \zeta \sum_i \ln \frac{\|x\|_1}{x_i} -
   \frac{1}{2} \zeta^2  \Bigl( \sum_i \frac{1}{x_i - \zeta} - \frac{m}{\|x\|_1
   / m - \zeta} \Bigr) . \]
This establishes the lemma. The coefficient of $\zeta^2$ above is $\geqslant
\; 0$ and equals 0 iff all elements of $x$ are equal ({\cite{HLP}}, Theorem
16).

\subsubsection*{Proof of Proposition \ref{prop:max}}

\begin{enumerate}
\item The proof is by contradiction. Assume that $u, v$ are two (distinct)
global maximizers of $G$ over $\mathcal{C} (0)$. It is not possible that both
of them have the same sum $s$: under the condition $\sum_i x_i = s$, we have
$G(x) = s \ln s - \sum_i x_i \ln x_i$ by {\eqref{eq:G}}. But the Shannon
entropy extended to all $x \succcurlyeq 0$ is strictly concave, so $G(x)$ has
a unique global maximizer over the convex domain $\mathcal{C} (0) \cap \left\{
x \mid \sum_i x_i = s \right\}$.

Next let $u$ and $v$ have different sums. We will derive a condition
{\tmem{necessary}} for both $u$ and $v$ to maximize $G$ and show that it is
contradicted by the scaling property of $G$.  Under our assumption that $G(u)
= G (v) = G^{\ast}$, the concavity of $G$ implies that any point on the line
segment between $u$ and $v$ must yield the same value, $G^{\ast}$, of $G$.
Thus the function $f (\alpha) = G (\alpha u + (1 - \alpha) v)$, $\alpha \in
[0, 1]$, must be a constant for all $\alpha$. Therefore $f' (\alpha)$ must be
0 for all $\alpha \in (0, 1)$. Rather than $f' (\alpha)$, it is easier to deal
with the expression for $f'' (\alpha)$. The constancy of $f' (\alpha)$ implies
that we must have $f'' (\alpha) \equiv 0$:
\[ f'' (\alpha) \; = \frac{\left( \sum_i u_i - \sum_i v_i \right)^2}{\alpha
   \sum_i u_i + (1 - \alpha)  \sum_i v_i} - \sum_i \frac{(u_i - v_i)^2}{\alpha
   u_i + (1 - \alpha) v_i} . \]
We will consider the condition $f'' \left( 1 / 2 \right) = 0$, and set $u_i -
v_i = z_i, u_i + v_i = w_i$. Then we have
\[ f'' \left( 1 / 2 \right) \; = \; \frac{\left( \sum_i z_i \right)^2}{\sum_i
   w_i} - \sum_i \frac{z_i^2}{w_i} = 0. \]
Further setting $q_i = w_i / \sum_i w_i$, since $\sum_i w_i > 0$ the above
condition is equivalent to $\left( \sum_i z_i \right)^2 - \sum_i (z_i^2 / q_i)
= 0$. But the l.h.s. is a strictly concave function of $q$, hence over the
convex set $q > 0, \sum_i q_i = 1$ it attains its global maximum of 0 at a
unique point $\hat{q}$, where $\widehat{q_i} = z_i / \sum_j z_j$.

So we have shown that $f'' \left( 1 / 2 \right) = 0 \Rightarrow w_i / \sum_j
w_j = z_i / \sum_j z_j$ for all $i$. This is equivalent to
\begin{equation}
  \forall i, \quad \frac{u_i + v_i}{\sum_j u_j + \sum_j v_j} = \frac{u_i -
  v_i}{\sum_j u_j + \sum_j v_j} \quad \tmop{or} \quad \; \frac{u_i}{v_i} =
  \frac{\sum_j u_j}{\sum_j v_j} . \label{eq:prop}
\end{equation}
This condition is \tmtextit{necessary} for $f' (\alpha)$ to be constant, in
particular 0, hence for $f (\alpha)$ to be constant. Finally, we can assume
w.l.o.g. that $u$ and $v$ are such that $\sum_i u_i > \sum_i v_i$, and then
{\eqref{eq:prop}} implies that there is some $c > 1$ s.t. $u = cv$. But then
the scaling property \ref{Gpropsc} in {\textsection}\ref{sec:Gbasic} says that
$G (u) = cG (v) > G (v)$, contradicting our initial assumption that both $u$
and $v$ maximize $G$.

\item If $\hat{x}$ were such a point, we would have $G (\hat{x}) > G
(x^{\ast})$ by Proposition \ref{prop:inc}, and this would contradict that
  $x^{\ast}$ is the global maximum.
\end{enumerate}

\subsubsection*{Proof of Proposition \ref{prop:theta_inf}}

Consider the equality constraints first. Writing them as $|A^E y - b^E |
\leqslant \delta |b^E |$, we see that this will be satisfied if $\max_i  |A^E
y - b^E |_i \leqslant \delta \min_i  |b_i^E |$, or $\|A^E y - b^E \|_{\infty}
\leqslant \delta |b^E |_{\min}$. Now for any $y \in \mathbb{R}^m$, $A^E y -
b^E = A^E  (y - x)$, since $x \in \mathcal{C} (0)$. Thus $\|A^E y - b^E
\|_{\infty} = \|A^E (y - x)\|_{\infty}$. But $\|A^E (y - x)\|_{\infty}
\leqslant \interleave A^E \interleave_{\infty}  \|y - x\|_{\infty}$, where the
(rectangular) matrix norm $\interleave \cdot \interleave_{\infty}$ is defined
as the largest of the $\ell_1$ norms of the rows\footnote{For any rectangular
matrix $A$ and compatible vector $x$, $\|Ax\|_{\infty} \leqslant \interleave A
\interleave_{\infty}  \|x\|_{\infty}$ holds because the l.h.s. is $\max_i 
|A_{i.} x|$. This is $\leqslant \max_i \sum_j |a_{i j} x_j | \leqslant \max_i
\|x\|_{\infty}  \|A_{i.} \|_1 = \|x\|_{\infty}  \interleave A
\interleave_{\infty}$.}. Therefore, to ensure $\|A^E y - b^E \|_{\infty}
\leqslant \delta |b^E |_{\min}$ it suffices to require that $\|y -
x\|_{\infty} \leqslant \delta |b^E |_{\min} / \interleave A^E
\interleave_{\infty}$, as claimed.

Turning to the inequality constraints, write them as $A^I  (x + y - x)
\leqslant b^I + \delta |b^I |$, or $A^I x - b^I \leqslant A^I  (x - y) +
\delta |b^I |$. Since $A^I x - b^I \leqslant 0$, this inequality will be
satisfied if $A^I  (y - x) \leqslant \delta |b^I |$. This will certainly hold
if $\max_i  (A^I  (y - x))_i \leqslant \delta \min_i  |b^I_i |$, which is
equivalent to $\|A^I (y - x)\|_{\infty} \leqslant \delta |b^I |_{\min}$. In
turn, this will hold if we require $\interleave A^I \interleave_{\infty}  \|y
- x\|_{\infty} \leqslant |b^I |_{\min}$.

For both types of constraints the final condition is stronger than necessary,
but more so in the case of inequalities. Finally, part 2 of the proposition
follows from part 1 since $\| [x] - x \|_{\infty} \leqslant 1 / 2$.

\subsubsection*{Proof of Proposition \ref{prop:x*scale}}

From (\ref{eq:x*j}) we can write the elements of $x^{\ast}$ in the form
$x^{\ast}_j = \left( \sum_i x^{\ast}_i \right) \mathcal{E}_j,$where
$\mathcal{E}_j$ is an expression involving the vectors $\lambda^E,
\lambda^{\tmop{BI}}$ and the matrices $A^E, A^{\tmop{BI}}$. The elements of
$\lambda^E, \lambda^{\tmop{BI}}$ are determined by substituting the
$x_j^{\ast}$ into the constraints. Thus the $k$th equality constraint leads to
an equation of the form
\begin{equation}
  \Bigl( \sum_i x^{\ast}_i \Bigr)  (\text{expression involving the
  $\mathcal{E}_j$}) \; = \; b^E_k \label{eq:constraint_scaling}
\end{equation}
and similarly for each binding inequality constraint. But the solution of a
system of equations of the form (\ref{eq:constraint_scaling}) is unchanged if
the $x^{\ast}_i$ on the l.h.s. and the $b^E$ and $b^{\tmop{BI}}$ on the r.h.s.
are both multiplied by the same constant $c > 0$.  This establishes the first
claim.  The claim about the maximum of $G$ follows from property \ref{Gpropsc}
of $G$ in the list of {\textsection}\ref{sec:G}.

Coming to the bounds on $x_1 + \cdots + x_m$, the fact that they scale with
$b$ is just a property of general linear programs. That is, if $y$ is the
solution to the linear program $\min_{x \in \mathbb{R}^m}  \sum_i \alpha_i
x_i$ subject to $Ax \leqslant b$, then $cy$ is the solution to $\min_{x \in
\mathbb{R}^m}  \sum_i \alpha_i x_i$ subject to $Ax \leqslant cb$. Similarly
for the maximum.

\subsubsection*{Proof of Proposition \ref{prop:s1s2bounds}}

For part 1, given $A^E x = b^E$, we have $\|A^E x\|_1 = \|b^E \|_1$. Now,
omitting the superscript to simplify the notation,
\begin{eqnarray*}
  \|Ax\|_1 & = & |a_{11} x_1 + \cdots + a_{1 m} x_m | + |a_{21} x_1 + \cdots +
  a_{2 m} x_m | + \cdots + |a_{\ell 1} x_1 + \cdots + a_{\ell m} x_m |\\
  & \leqslant & |a_{11} + a_{21} + \cdots + a_{\ell 1} |  |x_1 | + |a_{12} +
  a_{22} + \cdots + a_{\ell 2} |  |x_2 | + \cdots\\
  & \leqslant & \interleave A^T \interleave_{\infty}  \|x\|_1 .
\end{eqnarray*}
Hence $\interleave (A^E)^T \interleave_{\infty}  \|x\|_1 \geqslant \|b^E
\|_1$, and since $x \geqslant 0$, $\|x\|_1$ is simply the sum of the $x_i$.

For part 2, any $x \in \mathbb{R}^m$ satisfying $A^E x = b^E, A^I x \leqslant
b^I$ will satisfy $A^E x \leqslant b^E, A^I x \leqslant b^I$ as well. Divide
each inequality in this system by the smallest non-0 element of the l.h.s., if
that element is $< 1$, otherwise leave the inequality as is. Since each $x_i$
appears in some constraint, if we add all the above inequalities by sides the
resulting l.h.s. will be $\geqslant x_1 + \cdots + x_m$, and the r.h.s. will
be $\sum_i b^E_i / \alpha^E_i + \sum_i b^I_i / \alpha^I_i$, where the
$\alpha_i$ are defined as in the Proposition.

\subsubsection*{Proof of Proposition \ref{prop:nu*}}

First, the adjustment performed on $\tilde{\nu}$ is always possible: if $d < 0$
there must be at least $| d |$ elements of $n^{\ast} \chi^{\ast}$ that were
rounded to their floors, and if $d > 0$ to their ceilings. It is clear that the
adjustment makes $\nu^{\ast}$ sum to $n^{\ast}$. Now suppose that $k \in
\mathbb{N}$ and $\chi$ is an $m$-element density vector; then $k \chi$ sums to
$k$, and the sum of the rounded version $[k \chi]$ differs by no more than $m /
2$ from $k$. Thus $d \leqslant m / 2$.

For the bound on $\| \nu^{\ast} - x^{\ast} \|_1$, we first show that $\|
\nu^{\ast} - n^{\ast} \chi^{\ast} \|_1 \leqslant 3 m / 4$. The adjustment of
$\tilde{\nu}$ causes $d$ of the elements\tmtextsf{} of $\nu^{\ast}$ to differ
from the corresponding elements of $n^{\ast} \chi^{\ast}$ by $< 1$, and the
rest to differ by $\leqslant 1 / 2$, so $\| \nu^{\ast} - n^{\ast} \chi^{\ast}
\|_1 \leqslant \max_d  (d + (m - d) / 2) \leqslant 3 m / 4$. Next,
\[ \| \nu^{\ast} - x^{\ast} \|_1 = \| \nu^{\ast} - s^{\ast} \chi^{\ast} \|_1
   \leqslant \| \nu^{\ast} - n^{\ast} \chi^{\ast} \|_1 + \| n^{\ast}
   \chi^{\ast} - s^{\ast} \chi^{\ast} \|_1 \leqslant 3 m / 4 + | n^{\ast} -
   s^{\ast} |, \]
since $\chi^{\ast}$ sums to 1, and lastly $| n^{\ast} - s^{\ast} | < 1$ by
(\ref{eq:n*s*}).

That $\| \nu^{\ast} - x^{\ast} \|_{\infty} \leqslant 1$ follows from this
last statement and the fact that $\nu^{\ast}$ sums to $n^{\ast}$. Finally, the
bound on $\| f^{\ast} - \chi^{\ast} \|_1$ follows from that on $\| \nu^{\ast}
- n^{\ast} \chi^{\ast} \|_1$.

\subsubsection*{Proof of Lemma \ref{le:far}}

For brevity, in this proof we denote $G^{\ast} (0), x^{\ast} (0), \chi^{\ast}
(0)$ simply by $G^{\ast}, x^{\ast}, \chi^{\ast}$.

Given the vector $x^{\ast}$, set $s^{\ast} = x^{\ast}_1 + \cdots +
x^{\ast}_m$. Then from (\ref{eq:x*j}), $x^{\ast}_j = s^{\ast} e^{- (\lambda^E
\cdot A^E_{. j} + \lambda^{\tmop{BI}} \cdot A^{\tmop{BI}}_{. j})}$. Therefore
\begin{align*}
     \sum_i x^{\ast}_i \ln x^{\ast}_i & = \sum_i x^{\ast}_i  \bigl(\ln s^{\ast} -
     (\lambda^E \cdot A^E_{. i} + \lambda^{\tmop{BI}} \cdot A^{\tmop{BI}}_{.i})\bigr) \\
     & =  s^{\ast} \ln s^{\ast} - \sum_i x^{\ast}_i  (\lambda^E \cdot
     A^E_{. i} + \lambda^{\tmop{BI}} \cdot A^{\tmop{BI}}_{. i}) \\
     & = s^{\ast} \ln s^{\ast} - (\lambda^E \cdot b^E + \lambda^{\tmop{BI}}
     \cdot b^{\tmop{BI}}),
\end{align*}
since $x^{\ast}$ satisfies the equalities and the binding inequalities.
Substituting the above in (\ref{eq:G}), the maximum generalized entropy can be
expressed in terms of the Lagrange multipliers and the data as
\begin{equation}
  G^{\ast} = \lambda^E \cdot b^E + \lambda^{\tmop{BI}} \cdot b^{\tmop{BI}} .
  \label{eq:G*lambda}
\end{equation}
This implies that the quantity $\Lambda^{\ast}$ is at least as large as
$G^{\ast}$, as claimed.

Now if $\sigma$ is an {\tmem{arbitrary}} sequence with count vector $\nu$, its
probability under $\chi^{\ast}$ is
\[ {\Pr}_{\chi^{\ast}} (\sigma) \; = \; (\chi^{\ast}_1)^{\nu_1} \cdots
   (\chi^{\ast}_m)^{\nu_m} \]
where $\chi_j^{\ast} = e^{- (\lambda^E \cdot A^E_{. j} + \lambda^{\tmop{BI}}
\cdot A^{\tmop{BI}}_{. j})}$. Therefore
\begin{equation}
  {\Pr}_{\chi^{\ast}} (\sigma) = e^{- \xi (\nu)}, \qquad \text{where} \quad \xi
  (\nu) \; = \; \sum_i \lambda^E_i  (A^E_{i .} \cdot \nu) + \sum_i
  \lambda^{\tmop{BI}}_i  (A^{\tmop{BI}}_{i .} \cdot \nu) . \label{eq:xi}
\end{equation}
The rest of the proof is analogous to that of Proposition II.2 in
{\cite{entc2016}}. If $\nu$ is in $\mathcal{C} (\delta)$, then
\[ b^E_i - \delta | \beta^E_i | \leqslant A^E_{i .} \cdot \nu \leqslant b^E_i
   + \delta | \beta^E_i |, \qquad A^{\tmop{BI}}_{i .} \cdot \nu \leqslant
   b^{\tmop{BI}}_i + \delta | \beta^{\tmop{BI}}_i | . \]
Therefore from (\ref{eq:xi}), noting that $\lambda^{\tmop{BI}} \geqslant 0$
but the $\lambda_i^E$ can be positive or negative,
\[ \begin{array}{lll}
     \max_{\nu \in \mathcal{C} (\delta)} \xi (\nu) & \leqslant & \lambda^E
     \cdot b^E + (| \lambda^E | \cdot \nobracket | \beta^E |) \delta +
     \lambda^{\tmop{BI}} \cdot (b^{\tmop{BI}} + \delta | \beta^{\tmop{BI}}
     |),\\
     \min_{\nu \in \mathcal{C} (\delta)} \xi (\nu) & \geqslant & \lambda^E
     \cdot b^E - (| \lambda^E | \cdot | \beta^E |) \delta + \min_{\nu \in
     \mathcal{C} (\delta)} \sum_i \lambda_i^{\tmop{BI}}  (A^{\tmop{BI}}_{i .}
     \cdot \nu) .
   \end{array} \]
(The $| \cdot |$ around $\lambda^E$ cannot be removed.) Using
(\ref{eq:G*lambda}) in the above,
\begin{equation}
  \begin{array}{lll}
    \max_{\nu \in \mathcal{C} (\delta)} \xi (\nu) & \leqslant & G^{\ast} + (|
    \lambda^E | \cdot | \beta^E | + \lambda^{\tmop{BI}} \cdot |
    \beta^{\tmop{BI}} |) \delta,\\
    \min_{\nu \in \mathcal{C} (\delta)} \xi (\nu) & \geqslant & G^{\ast} - (|
    \lambda^E | \cdot | \beta^E |) \delta + \min_{\nu \in \mathcal{C}
    (\delta)} \sum_i \lambda_i^{\tmop{BI}}  (A^{\tmop{BI}}_{i .} \cdot \nu -
    b^{\tmop{BI}}_i)\\
    & = & G^{\ast} - (| \lambda^E | \cdot | \beta^E |) \delta - \Delta
    (\mathcal{C}(\delta)),
  \end{array} \label{eq:xi1}
\end{equation}
where
\[ \Delta (\mathcal{C}(\delta)) \; \triangleq \max_{\nu \in \mathcal{C}
   (\delta)}  \sum_i \lambda_i^{\tmop{BI}}  (b^{\tmop{BI}}_i -
   A^{\tmop{BI}}_{i .} \cdot \nu) . \]
Finally, for any p.d. $p$ and {\tmem{any}} $n$-sequence $\sigma$ with count
vector $\nu$, $\Pr_p (\sigma)$ is given by the expression in property
\ref{Gpropp} in {\textsection}\ref{sec:Gbasic}. Comparing that with
(\ref{eq:xi}), $\xi (\nu) = G (\nu) + nD (f\| \chi^{\ast})$, so by using
(\ref{eq:xi1})
\[ G^{\ast} - (| \lambda^E | \cdot \nobracket | \beta^E |) \delta - \Delta
   (\mathcal{C}(\delta)) \; \leqslant \; G (\nu) + nD (f \|
   \chi^{\ast}) \; \leqslant \; G^{\ast} + (| \lambda^E | \cdot | \beta^E | +
   \lambda^{\tmop{BI}} \cdot | \beta^{\tmop{BI}} |) \delta, \]
where $f = \nu / n$, and the claim of the lemma follows.

\subsubsection*{Proof of inequality (\ref{eq:nBn2})\label{app:pineq1}}

Let $y = \sqrt{n / 2}$. The sum $\sum_{k = 1}^m \binom{m}{k} y^k / \Gamma (k /
2)$ can be found in closed form by noticing that if it is split over even and
odd $k$, each of the sums is hypergeometric. However, the resulting expression
is too complicated for our purposes. We will obtain a tractable bound that
matches the highest power of $y$ in the sum, i.e. $y^m / \Gamma (m / 2)$.

We need an auxiliary fact, relating $\Gamma (k / 2)$ for $k < m$ to $\Gamma (m /
2)$. From Gautschi's inequality for the gamma function (see {\cite{HMF2}},
5.6.4) it follows that $\Gamma ((\mu - 1) / 2) > \Gamma (\mu / 2) / \sqrt{\mu /
  2}$, for any $\mu > 1$. Applying this recursively we find that for $k
\geqslant 1$
\begin{equation}
  \begin{aligned}
    \Gamma \Bigl( \frac{m-k}{2} \Bigr) & > \frac{2^{k/2}
      \Gamma(m/2)}{\bigl(m (m-1) \cdots (m-k+1) \bigr)^{1/2}} \\
    & > \frac{2^{k/2} \Gamma (m/2)}{m^{k/2} e^{- k (k - 1) / (4 m)}}
  \end{aligned} \label{eq:gammaineq}
\end{equation}
where the 2nd line follows by using $1 - z < e^{- z}$, for $z < 1$, in the
denominator of the first line.

Now pulling out the last term of our sum, reversing the order of the other
terms, and applying (\ref{eq:gammaineq}) to each term, we get
\begin{align*}
     \sum_{k = 1}^m \binom{m}{k} y^k  \frac{1}{\Gamma (k / 2)} & < 
     \frac{y^m}{\Gamma (m / 2)} + \frac{1}{\Gamma (m / 2)}  \sum_{k = 1}^{m -
     1} \binom{m}{m - k}  \left( \frac{m}{2} \right)^{k / 2} y^{m - k} e^{-
       \frac{k (k - 1)}{4 m}} \\
     & = \frac{y^m}{\Gamma (m / 2)} + \frac{y^m}{\Gamma (m / 2)}  \sum_{k =
     1}^{m - 1} \binom{m}{k}  \left( \frac{m}{2 y^2} e^{- \frac{k - 1}{2 m}}
     \right)^{k / 2} \\
     & < \frac{y^m}{\Gamma (m / 2)} + \frac{y^m}{\Gamma (m / 2)} \Bigl(
     \bigl( 1 + \sqrt{m/(2y^2)} \bigr)^m - 1 \Bigr) \\
     & = \frac{(n/2)^{\frac{m}{2}}}{\Gamma (m/2)} \bigl(1 + \sqrt{m/n}\bigr)^m,
\end{align*}
where in going from the 2nd to the 3d line we ignored the exponential factor and
the last term in the expansion of $\bigl(1 + \sqrt{m / (2 y^2)} \bigr)^m$, and
in the last line we substituted $y = \sqrt{n / 2}$.  The ratio of the sum and
this last expression tends to 1 as $n \rightarrow \infty$.

\subsubsection*{Proofs of inequality (\ref{eq:chatub})}

The first term is an upper bound on $c_1$. (\ref{eq:cref2}) is an inequality
of the type $x \geqslant \alpha \ln x + \beta$, with $\alpha, \beta \geqslant
0$. We will show that if $\alpha + \beta \geqslant 1$, this inequality is
satisfied by $x = 2 \alpha \ln (\alpha + \beta) + \beta$. [This expression is
motivated by the method of successive substitutions: with $x_0 = \beta$, we
get $x_2 = \alpha \ln (a \ln \beta + \beta) + \beta$; but this satisfies the
inequality only if $\beta < 1$.] Substituting into the inequality we get
\[ (\alpha + \beta)^2 \geqslant 2 \alpha \ln (\alpha + \beta) + \beta \quad
   \Leftrightarrow \quad \alpha + \beta \geqslant \frac{\alpha}{\alpha +
   \beta} 2 \ln (\alpha + \beta) + \frac{\beta}{\alpha + \beta} . \]
Therefore this will hold if $\alpha + \beta \geqslant \max (2 \ln (\alpha +
\beta), 1)$. Now we have assumed that $\alpha + \beta \geqslant 1$, and $x
\geqslant 2 \ln x$ is always true for $x > 0$, so our claim is established.
Turning to the case $\alpha + \beta < 1$, we can suppose that $\alpha < 1$,
otherwise we fall into the case $\alpha + \beta \geqslant 1$. Then it suffices
to find a $x$ that satisfies $x \geqslant \ln x + \beta$, and that is so for
$x = 1.5 \beta + \ln \beta$.

The third term in (\ref{eq:chatub}) is an upper bound on $c_3$. Write
(\ref{eq:cref4}) as $\frac{1}{2} \Sigma_1 - c \Sigma_2 \leqslant c^2 \eta
G^{\ast}$. This will hold if $c \geqslant \bigl( \sqrt{\Sigma_2^2 + 2 \Sigma_1
\eta G^{\ast}} - \Sigma_2 \bigr) / (2 \eta G^{\ast})$, so the r.h.s. can be
taken to be $c_3$. If $a, b \geqslant 0$, which is guaranteed by our
assumption that $x^{\ast} >\tmmathbf{1}$, then $\sqrt{a + b} < \sqrt{a} +
\sqrt{b}$, so $\sqrt{2 \Sigma_1 \eta G^{\ast}} / (2 \eta G^{\ast}) =
\sqrt{\frac{\sum_{i = 1}^m 1 / (x^{\ast}_i - 1)}{2 \eta G^{\ast}}}$ is an
upper bound on $c_3$.

\subsubsection*{Proof of Proposition \ref{prop:far}}

To ease the notation, let $\| x \| = s, \| y \| = t$. First we show that
\begin{equation}
  \left\| \frac{x}{s} - \frac{y}{t} \right\| \leqslant \vartheta \quad
  \Rightarrow \quad | \| x - y \| - | s - t | | \leqslant \min (s, t)
  \vartheta . \label{eq:farA1}
\end{equation}
We have
\begin{align*}
     \left\| \frac{x}{s} - \frac{y}{t} \right\| \leqslant \vartheta &
     \Leftrightarrow  \left\| \left( \frac{x}{s} - \frac{y}{s} \right) +
     \left( \frac{y}{s} - \frac{y}{t} \right) \right\| \leqslant \vartheta &
     \Rightarrow  \left| \left\| \frac{x}{s} - \frac{y}{s} \right\| - \left\|
     \frac{y}{s} - \frac{y}{t} \right\| \right| \leqslant \vartheta \\
     & \Leftrightarrow  \left| \frac{1}{s}  \| x - y \| - \left\| \left(
     \frac{1}{s} - \frac{1}{t} \right) y \right\| \right| \leqslant
     \vartheta & \Leftrightarrow  \left| \frac{1}{s}  \| x - y \| - \left|
     \frac{1}{s} - \frac{1}{t} \right| t \right| \leqslant \vartheta \\
     & \Leftrightarrow | \| x - y \| - | s - t | | \leqslant s \vartheta .
     &
\end{align*}
Exchanging $x$ with $y$ and $s$ with $t$ in this derivation, it also follows
that
\[ \left\| \frac{x}{s} - \frac{y}{t} \right\| \leqslant \vartheta \quad
   \Rightarrow \quad | \| x - y \| - | s - t | | \leqslant t \vartheta, \]
and this establishes (\ref{eq:farA1}). Now (\ref{eq:farA1}) implies that
\[ \left\| \frac{x}{\| x \|} - \frac{y}{\| y \|} \right\| \leqslant
   \vartheta \quad \Rightarrow \quad \| x - y \| \; \leqslant \; \min (\| x
   \|, \| y \|) \vartheta + | \| x \| - \| y \| | \]
and taking the contrapositive of this
\[ \| x - y \| \; > \; \min (\| x \|, \| y \|) \vartheta + | \| x \| - \| y \|
   | \quad \Rightarrow \quad \left\| \frac{x}{\| x \|} - \frac{y}{\| y \|}
   \right\| > \vartheta, \]
from which the claim of the proposition follows.

\subsubsection*{Proof of the inequality in (\ref{eq:numBn1n2bound})}

This is an improvement over bounding the sum in the second line of
(\ref{eq:numBn1n2bound}) by simply pulling out $e^{- \matheuler^{\ast}
\vartheta^2 n_1}$ and then bounding the rest by an integral. Splitting the sum
around the point $n^{\ast}$,
\begin{align*}
  \sum^{n_2}_{n = n_1} \frac{\left( \sqrt{n} + \sqrt{m}
    \right)^m}{\sqrt{n}} e^{- \gamma^{\ast} \vartheta^2 n} & \leqslant & \\
  e^{- \gamma^{\ast} \vartheta^2 n_1}  \sum_{n = n_1}^{n^{\ast}}
  \frac{\left( \sqrt{n} + \sqrt{m} \right)^m}{\sqrt{n}} & + e^{-
    \gamma^{\ast} \vartheta^2  (n^{\ast} + 1)}  \sum_{n = n^{\ast} + 1}^{n_2}
  \frac{\left( \sqrt{n} + \sqrt{m} \right)^m}{\sqrt{n}} \quad \leqslant\\
  2 e^{- \gamma^{\ast} \vartheta^2 s_1}  \int_{\sqrt{s_1}}^{\sqrt{n^{\ast}
      + 1}} \left( u + \sqrt{m} \right)^m d u \; & + 2 e^{- \gamma^{\ast}
    \vartheta^2  (s^{\ast} + 1)}  \int_{\sqrt{n^{\ast} + 1}}^{\sqrt{s_2 + 2}}
  \left( u + \sqrt{m} \right)^m d u,
\end{align*}
since the summand is an increasing function of $n$. The last line can be
written as
\begin{align*}
     & \frac{2}{m + 1}  \left( \sqrt{n^{\ast} + 1} + \sqrt{m} \right)^{m + 1} 
     (e^{- \gamma^{\ast} \vartheta^2 s_1} - e^{- \gamma^{\ast} \vartheta^2 
     (s^{\ast} + 1)}) + \\
     & \frac{2}{m + 1}  \left( \left( \sqrt{s_2 + 2} + \sqrt{m} \right)^{m + 1}
     e^{- \gamma^{\ast} \vartheta^2  (s^{\ast} + 1)} - \left( \sqrt{s_1} +
     \sqrt{m} \right)^{m + 1} e^{- \gamma^{\ast} \vartheta^2 s_1} \right)
\end{align*}
and the desired result follows by neglecting the second exponential from each
of the two summands.

\subsubsection*{Proof of Theorem \ref{th:cdist3}}

We minimize the max of $c_2 (\delta), c_3 (\delta)$ by setting them equal to
each other. Substituting $c_2$ for $c$ into (\ref{eq:nlb3}) which defines
$c_3$, we get the equation for $\delta$ in the theorem:
\begin{equation}
  \frac{2 \gamma^{\ast} \vartheta^2 s_1}{\vartheta_{\infty}}  \frac{1}{\delta}
  + m \ln \delta \; = \; \ln \frac{C''_3}{\varepsilon B'} +
  \frac{\Lambda^{\ast}}{\vartheta_{\infty}} - m \ln \vartheta_{\infty} .
  \label{eq:fdelta}
\end{equation}
Let $f (\delta)$ stand for the function of $\delta$ on the l.h.s. This function
decreases for $\delta < 2 \gamma^{\ast} \vartheta^2 s_1 / \allowbreak (m
\vartheta_{\infty})$. From the condition between $\vartheta$ and $\delta$ of
Theorem \ref{th:cdist2}, we must have $\delta < \delta_{\max} = 2 \gamma^{\ast}
\vartheta^2 s_1 / \Lambda^{\ast}$. So if $2 \gamma^{\ast} \vartheta^2 s_1 / (m
\vartheta_{\infty}) > \delta_{\max}$, which will hold if $\Lambda^{\ast}
\geqslant m \vartheta_{\infty}$, then $f (\delta)$ will decrease with $\delta
\in (0, \delta_{\max})$. If $f (\delta_{\max})$ is less than the r.h.s. of
(\ref{eq:fdelta}) then (\ref{eq:fdelta}) will have a root $\delta_0 \in (0,
\delta_{\max}]$. This condition on $f (\delta_{\max})$ boils down to
\begin{equation}
  \left( \frac{2 \gamma^{\ast} \vartheta_{\infty} \vartheta^2
  s_1}{\Lambda^{\ast}} \right)^m < \; \frac{C''_3}{\varepsilon B'}, \quad
  \text{or} \quad \varepsilon^{1 / m} \vartheta^2 \; < \;
  \frac{\Lambda^{\ast}}{2 \gamma^{\ast} \vartheta_{\infty} s_1}  \left(
  \frac{C''_3}{B'} \right)^{1 / m} . \label{eq:fdmax}
\end{equation}
To arrive at the condition of the theorem we find a simple lower bound on
$(C''_3 / B')^{1 / m}$. From (\ref{eq:C3pp}),
\[ C''_3 \; > \; \left( \sqrt{s^{\ast} + 2} + \sqrt{m} \right)^{m + 1} \; > \;
   \left( \sqrt{s^{\ast} + m + 2} \right)^{m + 1} . \]
Therefore from (\ref{eq:nlb2})
\begin{align*}
     \frac{C''_3}{B'} & > \frac{2 \pi^{m / 2} e^{m / 12}  \left(
     \sqrt{s^{\ast} + m + 2} \right)^{m + 1}}{(m + 1) \Gamma (m / 2) 
     \sqrt{s^{\ast}}} \\
     & > \frac{2 \pi^{m / 2} e^{m / 12}}{(m + 1) \Gamma (m / 2)}  \left(
     \sqrt{s^{\ast} + m + 2} \right)^m
\end{align*}
where in the first line we used the fact that the product of the last two
factors in the expression (\ref{eq:nlb2}) for $B'$ is $< \; 1$. It follows
that
\begin{align*}
     \left( \frac{C''_3}{B'} \right)^{1 / m} & > \frac{2^{1 / m}  \sqrt{\pi}
     e^{1 / 12}}{((m + 1) \Gamma (m / 2))^{1 / m}} \sqrt{s^{\ast} + m + 2} \\
     & = \frac{2^{1 / m}  \sqrt{\pi} e^{1 / 12}  \sqrt{m}}{((m + 1) \Gamma
     (m / 2))^{1 / m}}  \sqrt{s^{\ast} / m + 1 + 2 / m} \\
     & > 2 \sqrt{s^{\ast} / m + 1} .
\end{align*}
[To go from the 2nd to the 3d line, it can be shown that the first factor on
the 2nd line is an increasing function of $m$; its minimum occurs at $m = 2$
and is $\approx \; 2.22$.] It follows that condition (\ref{eq:fdmax}) for the
existence of the root $\delta_0$ will be satisfied if
\[ \varepsilon^{1 / m} \vartheta^2 \; \leqslant \;
   \frac{\Lambda^{\ast}}{\gamma^{\ast} \vartheta_{\infty} s_1}  \sqrt{s^{\ast}
   / m + 1}, \]
as stated in the theorem. Now since we have ensured $c_2 (\delta_0) = c_3
(\delta_0)$, we can take $\hat{c} = \max \bigl(c_2 (\delta_0), c_1 \bigr)$,
where $c_1$ is as in Theorem \ref{th:cdist2}.

Finally, it is quite likely that $c_2 (\delta_0) > c_1$ so that $\hat{c} = c_2
(\delta_0)$. Given that $\delta_0 < \delta_{\max} = 2 \gamma^{\ast}
\vartheta^2 s_1 / \Lambda^{\ast}$ and $\Lambda^{\ast} \geqslant m
\vartheta_{\infty}$, it can be seen that this will be so if if $\vartheta <
s^{\ast} / (2 \gamma^{\ast} s_1)$.

\end{document}